\documentclass[english,journal,letterpaper]{IEEEtran}
\usepackage[T1]{fontenc}
\usepackage[latin9]{inputenc}
\usepackage{amsmath}
\usepackage{amssymb}
\usepackage{graphicx}

\makeatletter
\usepackage{cite}
\usepackage{times}   
\newtheorem{theorem}{Theorem}[section]
\newtheorem{definition}[theorem]{Definition}
\newtheorem{remark}[theorem]{Remark}

\newtheorem{lemma}[theorem]{Lemma}
\newtheorem{proposition}[theorem]{Proposition}
\newtheorem{aplemma}{Lemma}[section]

\usepackage[all]{xy}

\usepackage{balance}

\usepackage{tikz}
\usetikzlibrary{positioning, arrows, matrix}

\usepackage{psfrag}

\usepackage{dsfont}
\newcommand{\mathbbm}[1]{\mathds{#1}}

\newtheorem{assumption}{Assumption}[section]

\usepackage{url}

\mathchardef\mhyphen="2D

\renewcommand{\IEEEQED}{\IEEEQEDopen}

\newenvironment{nopf}{\comment
}{\endcomment
}

\newenvironment{seearxiv}{
}{
}

\newenvironment{inarxiv}{
}{
}

\usepackage{verbatim}

\usepackage{amsmath,etoolbox}
\makeatletter
\newtoggle{reducedop}
\newcommand{\reduceoperatorsize}[1]{%
  \csletcs{#1saved}{#1}%
  \csletcs{#1@saved}{#1@}%
  \@namedef{#1}{%
    \@ifstar
      {\togglefalse{reducedop}\@nameuse{#1saved}}%
      {\toggletrue{reducedop}\@nameuse{#1saved}}%
  }
  \@namedef{#1@}{%
    \iftoggle{reducedop}
      {\reduced@operator{#1}}
      {\@nameuse{#1@saved}}
  }%
}
\newcommand{\reduced@operator}[1]{%
  \mathop{\mathpalette\reduced@operator@i{#1}\relax}%
}
\newcommand\reduced@operator@i[2]{%
  \ifx#1\displaystyle\textstyle\else#1\fi
  \csname #2@saved\endcsname
}
\makeatother

\makeatother

\usepackage{babel}
\begin{document}

\title{\LARGE \bf Analysis of Stochastic Switched Systems with Application
to Networked Control Under Jamming Attacks}

\author{Ahmet Cetinkaya, Hideaki Ishii, and Tomohisa Hayakawa \thanks{A. Cetinkaya and H. Ishii are  with the Department of Computer Science, Tokyo Insitute of Technology, Yokohama, 226-8502, Japan. {\tt\small{ahmet@sc.dis.titech.ac.jp,ishii@c.titech.ac.jp}}}
\thanks{T. Hayakawa is with the Department of Systems and Control Engineering, Tokyo Institute of Technology, Tokyo 152-8552, Japan. {\tt\small{hayakawa@mei.titech.ac.jp}}}
\thanks{This work was supported  in part by the JST CREST Grant  No.\ JPMJCR15K3 and by JSPS under Grant-in-Aid for Scientific Research Grant No.~15H04020.}}

\maketitle
\begin{abstract} We investigate the stability problem for discrete-time
stochastic switched linear systems under the specific scenarios where
information about the switching patterns and the probability of switches
are not available. Our analysis focuses on the average number of times
each mode becomes active in the long run and, in particular, utilizes
their lower- and upper-bounds. This setup is motivated by cyber security
issues for networked control systems in the presence of packet losses
due to malicious jamming attacks where the attacker's strategy is
not known a priori. We derive a sufficient condition for almost sure
asymptotic stability of the switched systems which can be examined
by solving a linear programming problem. Our approach exploits the
dynamics of an equivalent system that describes the evolution of the
switched system's state at every few steps; the stability analysis
may become less conservative by increasing the step size. The computational
efficiency is further enhanced by exploiting the structure in the
stability analysis problem, and we introduce an alternative linear
programming problem that has fewer variables. We demonstrate the efficacy
of our results by analyzing networked control problems where communication
channels face random packet losses as well as jamming attacks. \end{abstract}


\section{Introduction }

In the recent studies of hybrid systems, switched systems represent
an important and fundamental class due to their simple structures.
Switched systems are composed of a number of subsystems that possess
different continuous dynamics and a discrete-valued switching mode
signal which determines the active subsystem. Complicated behaviors
in the state evolutions can be demonstrated, depending on the nature
of the switching as well as the dynamics of the subsystems. For this
reason, analyses of switched systems concerning their stability and
performance have posed various challenges and led researchers to interesting
results \cite{ajlp_survey_liberzon1999basic,ajlp_survey_switched_lin2009stability}.

Recently, the importance of this class of hybrid systems is rising
from the application side. In particular, in large-scale networked
control systems, switchings in the system dynamics frequently take
place whenever the status in the communication networks changes. In
such systems, network channels connect the plant having many sensors
and actuators with remote controllers. Thus, transmission times and
patterns of control-related signals affect the dynamics of the plant
as well as the overall closed-loop system. Furthermore, the communication
is often unreliable in that the transmitted data may become lost,
not reaching the destination depending on the condition of the channels.
Since such channel behaviors are commonly modeled in a probabilistic
manner, the study of networked control systems often call for the
framework of stochastic switched systems.

Stability of switched systems has been studied for different types
of mode signals. The case of (deterministic) arbitrary switching has
been explored in a number of works including \cite{ajlp_arbitrary_switching_yliberzon1999stability,ajlp_arbitrary_switching_daafouz2002_arbitrary_switching_lyapunov},
establishing conditions under which a switched system remains stable
for all possible mode switching scenarios. The switching frequency
in the mode signal can be restricted by utilizing dwell-time and average
dwell-time notions. The works \cite{ajlp_liberzon2003,hespanha1999stability}
dealt with systems composed of only stable subsystems while \cite{ajlp_average_unstable_zhai2001stability,ajlp_average_unstable_zhang2014stability}
made extensions to systems also consisting of unstable subsystems.
In addition, \cite{ajlp_ishii2005randomized} investigated the problem
of designing state-dependent switching rules to guarantee stability.

For stochastic mode signals, stability issues of switched systems
have attracted considerable attention as well (see \cite{ajlp_survey_stochastic_teel2014stability,ajlp_markov_survey_shi2015_survey}
and the references therein). In most cases, stability results for
such systems rely on statistical information on the mode signal such
as the probability of mode switches and the stationary distributions
associated with the modes. An important class there is that of Markov
jump systems, for which the mode signal is dominated by underlying
Markov chains \cite{fang1995stability,costa2004discrete,zhou2014adaptive,zhu2017razumikhin,wang2017stability}.
Moreover, some works characterized both stochastic and deterministic
effects in the switching of the mode signal, referred to as ``dual
switching'' \cite{ajlp_bolzern2010markov,ajlp_both_bolzern2016design}.

In this paper, we investigate a stability problem for discrete-time
stochastic switched linear systems with a special emphasis on the
case where information about the mode switching probabilities or the
stationary distributions are not available for analysis. Our interest
stems from the current research activities on cyber security of networked
control systems \cite{HSF-SM:ct-13,de2015inputtran,cetinkaya2016tac}.
Today, more control systems are connected to the Internet and wireless
networks for their remote operation and monitoring. Such communication
settings significantly increase the risk to be targeted by malicious
cyber attackers. Clearly, the system dynamics can change, for example,
if attackers interfere with the communication of the control related
signals. Under such conditions, the networked system may be represented
as a stochastic switched system, but the a priori knowledge on the
switching, whether it is deterministic or probabilistic, would be
extremely limited. In the networked control literature, recent works
dealt with denial-of-service (DoS) and jamming attacks \cite{xu2005feasibility,amin2009,pelechrinis2011}
or packet drops by compromised routers \cite{mizrak2009detecting,shu2015privacy}.

Our problem formulation is based on the approach in our recent work
\cite{cetinkaya2016tac}; see also \cite{ahmet-cdc,cetinkaya2015output,ahmetnecsys2016,ahmetifac2017worldcongress}
for more related results. There, we focused on the stochastic stability
of a networked control system under malicious jamming attacks, where
the jamming causes losses of data during packet transmissions due
to strong interferences. The normal operation and the dynamics under
jamming attacks are represented with two different modes. The transmission
failure instants and the probability of transmission failures are
influenced by the attacker's actions and hence not available for analysis.
In \cite{cetinkaya2016tac}, we introduced a novel stochastic model
regarding the timings of jamming using only the asymptotic tail probabilities
of average transmission failures. Roughly speaking, it corresponds
to the bound on the average ratio of time that the communication is
blocked due to jamming. A key property of the model lies in its generality,
being capable to describe both malicious and nonmalicious data losses.
Specifically, it can represent discrete-time versions of the jamming
and denial-of-service attack models in \cite{HSF-SM:ct-13,de2015inputtran}
as well as random packet loss models based on Bernoulli and Markov
processes \cite{hespanha2007,ishii2009,xie2009stability,okano2014}.

In this paper, as a generalization of this class of systems in networked
control under jamming attacks, we consider stochastic switched systems
having arbitrary number of modes. The limited knowledge on the model
signal is represented by the lower- and upper-bounds on the percentage
of the time each mode is active in the long run. We pay special attention
to improving the conservatism and computational efficiency for the
stability analysis. It is established that linear programming can
be used for identifying the worst mode switching scenario in terms
of stability, and we assess the stability of the system under that
scenario. This stability assessment approach is different from the
matrix inequality-type stability conditions obtained with the Lyapunov-based
stability analysis in \cite{cetinkaya2016tac}. Certain aspects are
generalized in the linear programming-based conditions and for this
reason less conservative analysis is possible. In this paper, we will
discuss these points in detail and also illustrate by numerical examples.
Moreover, the proposed approach differs from those in \cite{ajlp_ly_polanski1997lyapunov,ajlp_ly_liu2009stability,ajlp_ly_baier2012linear,ajlp_ly_zhu2014l1},
where linear programming is used for constructing Lyapunov functions.

In the course of our analysis, we first construct a new system that
describes the evolution of the switched system's state at every $h\in\mathbb{N}$
steps. For a given switched system with $M$ modes, this new ``lifted''
version of the system is composed of $M^{h}$ modes, each of which
is identified by a sequence of $h$ numbers indicating a progression
of modes in the original system. The advantage of the approach is
that the conservatism in the analysis may be reduced by increasing
the size of $h$. This is possible because, with large $h$, stability/instability
properties of more switching patterns are taken into account. Furthermore,
to reduce the computational complexity, we exploit the problem structure
and develop an alternative linear programming problem. This problem
shares the same optimal objective function value with the original
problem, but involves fewer variables, which are in fact polynomial
with respect to the problem size.

The idea of studying a new system that describes the state's evolution
at every few steps has previously been employed in \cite{bolzern2004almost,ajlp_huang2011set,lee2006uniform,ajlp_constrained_philippe2016stability_constrained}.
The works \cite{bolzern2004almost} and \cite{ajlp_huang2011set}
explore Markov jump systems, but assume the full knowledge of the
Markov chain. On the other hand, in \cite{lee2006uniform} and \cite{ajlp_constrained_philippe2016stability_constrained},
stability problems under constrained switching rules are addressed,
where all possible switching patterns are identified through graphs
that indicate the constraints in the switching. In our problem setting,
information about the possible switching patterns is not available,
and moreover, we consider scenarios where switching is allowed between
all pairs of modes. For this particular setting, the stability results
in \cite{lee2006uniform,ajlp_constrained_philippe2016stability_constrained}
require all subsystems to be stable. Therefore, they are not applicable
for our problem since we allow both stable and unstable subsystem
dynamics. In addition to the analysis of switched systems, an approach
similar to the investigation of the dynamics over multiple time steps
has also been used for establishing the convergence of risk-sensitive
and risk sensitive like filters in \cite{levy2016contraction,zorzi2015convergence}. 

We show in the paper that our results are applicable to a wide class
of mode signals. In particular, it includes those that are the outputs
of hidden Markov chains whose actual state space and transition probabilities
are unknown. In networked control problems under random and malicious
packet transmission failures such signals may arise in periodic attacks
discussed in \cite{HSF-SM:ct-13} as well as random packet losses
on channels described with Markov models studied in \cite{sadeghi2008,ellis2014}.
We apply our stability results in two networked control problems,
where the plant and the remotely located controller exchange information
packets over one or more channels subject to packet losses due to
random communication errors or malicious jamming attacks.

The rest of the paper is organized as follows. In Section~\ref{sec:Stochastic-Switched-System},
we explain the switched system dynamics and obtain sufficient stability
conditions by utilizing bounds on the average number of times each
mode is active in the long run. In Section~\ref{sec:Linear-Programming-Methods},
we present our approach for checking stability conditions by means
of solving linear programming problems. Moreover, in Section~\ref{sec:Application-to-Networked},
we discuss the application of our results in the problem of networked
control under jamming attacks. We present numerical examples in Section~\ref{sec:Numerical-Example}.
Finally, in Section~\ref{sec:Conclusion}, we conclude the paper.

We note that part of the results in Sections~\ref{sec:Stochastic-Switched-System}
and \ref{sec:Linear-Programming-Methods} appeared in our preliminary
report \cite{cetinkaya2016enhanced} for the special case of two modes
in the context of a networked control problem. Here, we provide complete
proofs and more detailed discussions for the general modes case.

We use a fairly standard notation in the paper. Specifically, $\mathbb{N}$
and $\mathbb{N}_{0}$ respectively denote positive and nonnegative
integers. A finite-length sequence of ordered elements $q_{1},q_{2},\ldots,q_{h}$
is represented by $q=(q_{1},q_{2},\ldots,q_{h})$. We use $\lfloor\cdot\rfloor$
to denote the largest integer that is smaller than or equal to its
real argument. The notation $\mathrm{\mathbb{P}}[\cdot]$ denotes
the probability on a probability space $(\Omega,\mathcal{F},\mathbb{P})$.
Furthermore, we utilize $\mathbbm{1}[E]:\Omega\to\{0,1\}$ for the
indicator of the event $E\in\mathcal{F}$, that is, $\mathbbm{1}[E](\omega)=1$
for $\omega\in E$, and $\mathbbm{1}[E](\omega)=0$ for $\omega\notin E$. 

\section{Stochastic Switched System Stability Analysis \label{sec:Stochastic-Switched-System}}

In this section we first describe the dynamics of a stochastic switched
system. We then discuss the stability problem and provide sufficient
almost sure asymptotic stability conditions. 

\subsection{Switched System Dynamics}

Consider the discrete-time switched linear system with $M\in\mathbb{N}$
modes described by 
\begin{align}
x(t+1) & =A_{r(t)}x(t),\quad x(0)=x_{0},r(0)=r_{0},t\in\mathbb{N}_{0},\label{eq:switched-system}
\end{align}
 where $x(t)\in\mathbb{R}^{n}$ denotes the state vector, $\{r(t)\in\{1,\ldots,M\}\}_{t\in\mathbb{N}_{0}}$
is the mode signal, and $A_{s}\in\mathbb{R}^{n\times n}$, $s\in\{1,\ldots,M\}$,
represent the system matrices for each mode. We use $\mathcal{M}\triangleq\{1,\ldots,M\}$
to denote the set of modes. 

The mode signal $\{r(t)\in\mathcal{M}\}_{t\in\mathbb{N}_{0}}$ is
assumed to be a stochastic process that satisfies the following assumption. 

\begin{assumption} \label{MainAssumption-1} There exist scalars
$\underline{\rho}_{s},\overline{\rho}_{s}\in[0,1]$, $s\in\mathcal{M}$,
such that 
\begin{align}
\liminf_{k\to\infty}\frac{1}{k}\sum_{t=0}^{k-1}\mathbbm{1}[r(t)=s] & \geq\underline{\rho}_{s},\label{eq:infcond}\\
\limsup_{k\to\infty}\frac{1}{k}\sum_{t=0}^{k-1}\mathbbm{1}[r(t)=s] & \leq\overline{\rho}_{s},\label{eq:supcond}
\end{align}
 almost surely. 

\end{assumption} \vskip 5pt

In Assumption~\ref{MainAssumption-1}, the scalars $\underline{\rho}_{s}$
and $\overline{\rho}_{s}$ respectively represent lower- and upper-bounds
on the long-run average number of times mode $s$ is active. If no
information is available on the long-run average for mode $s$, the
scalars $\underline{\rho}_{s}$ and $\overline{\rho}_{s}$ can be
selected as $\underline{\rho}_{s}=0$ and $\overline{\rho}_{s}=1$,
since (\ref{eq:infcond}) and (\ref{eq:supcond}) are trivially satisfied
with those values for all $s\in\mathcal{M}$. 

Our main motivation for considering Assumption~\ref{MainAssumption-1}
is to analyze switched systems for which precise information about
the mode switching rules is not available for analysis. In Section~\ref{sec:Stability-Analysis},
we show that the scalars $\underline{\rho}_{s}$, $\overline{\rho}_{s}$,
$s\in\mathcal{M}$, can be used for analysis, even if we do not know
how the mode may switch at each time $t$. 

Assumption \ref{MainAssumption-1} allows the mode signal $\{r(t)\in\mathcal{M}\}_{t\in\mathbb{N}_{0}}$
to be generated in many different ways either randomly according to
a probability distribution or in a deterministic fashion. For instance,
$\{r(t)\in\mathcal{M}\}_{t\in\mathbb{N}_{0}}$ may be a stationary
and ergodic stochastic process with stationary distribution $\pi\in[0,1]^{M}$.
In that case $\underline{\rho}_{s}$ and $\overline{\rho}_{s}$ would
be scalars that satisfy $\underline{\rho}_{s}\leq\pi_{s}\leq\overline{\rho}_{s}$,
$s\in\mathcal{M}$. On the other hand, $\{r(t)\in\mathcal{M}\}_{t\in\mathbb{N}_{0}}$
may also represent a deterministically generated switching sequence.
For example, a particular mode switching pattern of length $T\in\mathbb{N}$
can be repeated to create a periodic switching sequence. In that case,
$\underline{\rho}_{s}$ and $\overline{\rho}_{s}$ would correspond
to lower- and upper-bounds on the ratio of the number of times mode
$s$ is active in the $T$-length switching pattern. In the case where
$\{r(t)\in\mathcal{M}\}_{t\in\mathbb{N}_{0}}$ is deterministic, the
limits $\lim_{k\to\infty}\frac{1}{k}\sum_{t=0}^{k-1}\mathbbm{1}[r(t)=q]$,
$q\in\mathcal{M}$, when they exist, correspond to the ``discrete
event rate'' utilized by \cite{hassibi1999control,zhang2001stability}
for deterministic systems. 

\begin{remark} \label{RemarkNotMarkovChain} In the literature of
stochastic switched systems the mode signal is typically characterized
as a Markov chain \cite{costa2004discrete}. In this paper, $\{r(t)\in\mathcal{M}\}_{t\in\mathbb{N}_{0}}$
is not necessarily a Markov chain. In fact in certain cases, we may
have
\begin{align*}
\mathbb{P}[r(t+1)=q|r(0),r(1),\ldots,r(t)] & \neq\mathbb{P}[r(t+1)=q|r(t)],
\end{align*}
which indicates that $r(\cdot)$ fails to satisfy the Markov property
\cite{norris2009}. Furthermore, $r(\cdot)$ may also depend on other
processes. Our main motivation for considering this general setup
for the mode signal $\{r(t)\in\mathcal{M}\}_{t\in\mathbb{N}_{0}}$
comes from the networked control problem under jamming attacks. As
we illustrate in Section~\ref{sec:Application-to-Networked}, a networked
control system under jamming attacks can be represented by the switched
system (\ref{eq:switched-system}). The mode signal of such a switched
system is not necessarily a Markov process, since the attacker can
influence the mode signal in various ways. Consequently, the analysis
of the switched system (\ref{eq:switched-system}) requires methods
that are different from those utilized in switched systems with Markov
modes \cite{costa2004discrete}. \end{remark}

In our stability analysis in the following section, we utilize the
scalars $\underline{\rho}_{s}$ and $\overline{\rho}_{s}$ satisfying
(\ref{eq:infcond}) and (\ref{eq:supcond}) instead of transition
probabilities or switching patterns. In Section~\ref{sec:Application-to-Networked},
we model networked control systems as switched systems, where the
mode signals represent the state of certain communication channels
that face jamming. There, $\underline{\rho}_{s}$ and $\overline{\rho}_{s}$
are characterized based on the information about the average number
of times transmissions fail due to attacks in the long run. 

\subsection{Stability Analysis \label{sec:Stability-Analysis}}

In this section, we explore the stability of the switched system (\ref{eq:switched-system}),
where the mode signal satisfies Assumption~\ref{MainAssumption-1}.
We use the stochastic stability notion of \emph{almost sure asymptotic
stability} in our analysis. 

\begin{definition} \label{Definition-Almost-Sure-Stability} The
zero solution $x(t)\equiv0$ of the stochastic system (\ref{eq:switched-system})
is \emph{almost surely stable} if for each $\epsilon>0$ and $\bar{p}>0$,
there exists $\delta=\delta(\epsilon,\bar{p})>0$ such that if $\|x_{0}\|_{2}<\delta$,
then 
\begin{align}
\mathbb{P}[\sup_{t\in\mathbb{N}_{0}}\|x(t)\|_{2}>\epsilon] & <\bar{p},\label{eq:definition-stability}
\end{align}
where $\|\cdot\|_{2}$ denotes the Euclidean norm. Moreover, the zero
solution $x(t)\equiv0$ is \emph{asymptotically stable almost surely}
if it is almost surely stable and 
\begin{align}
\mathbb{P}[\lim_{t\to\infty}\|x(t)\|_{2}=0] & =1.\label{eq:definition-convergence}
\end{align}

\end{definition} \vskip 10pt

Stability of discrete-time switched systems has been investigated
in many studies under different assumptions on the mode signal. For
instance, in several works (see \cite{ajlp_survey_switched_lin2009stability}
and the references therein) researchers explore stability of systems
with a form similar to (\ref{eq:switched-system}) under arbitrary
switching. A necessary condition is stability of each mode (i.e.,
$A_{1},A_{2},\ldots,A_{M}$ need to be Schur matrices). In our problem
setting we allow some of the modes to be unstable, and hence this
approach is not applicable. 

On the other hand, researchers also studied stability of systems similar
to (\ref{eq:switched-system}) for the case where $r(\cdot)$ is a
Markov process (see, e.g., \cite{fang1995stability,bolzern2004almost,costa2004discrete}).
The stability analysis in those studies relies on transition probabilities
and stationary distributions associated with the Markov process that
characterize the switching sequence. Note again that in our case,
$r(\cdot)$ need not be a Markov process. Furthermore, to account
for the uncertainty in generation of the mode signal, we assume that
statistical information concerning transition probabilities and stationary
distributions is not available. Hence, the stability results reported
in the above-mentioned literature are not applicable to the present
problem. 

In our stability analysis of the switched system, we follow the approach
in \cite{bolzern2004almost,lee2006uniform} and investigate the evolution
of the system's state at every $h\in\mathbb{N}$ steps. First, let
$\mathcal{M}^{h}$ denote the set of sequences of length $h$ with
entries in $\mathcal{M}$, that is, 
\begin{align*}
\mathcal{M}^{h} & \triangleq\{(q_{1},q_{2},\ldots,q_{h})\colon q_{j}\in\mathcal{M},j\in\{1,\ldots,h\}\}.
\end{align*}
With this definition, $q_{i}$ ($i$th entry of a sequence $q$) represents
a mode in the set of modes $\mathcal{M}$. Now, let $\{\bar{r}(i)\in\mathcal{M}^{h}\}_{i\in\mathbb{N}_{0}}$
be a \emph{sequence-valued} process defined by 
\begin{align}
\bar{r}(i) & \triangleq(r(ih),r(ih+1),\ldots,r((i+1)h-1)),\,\,\,i\in\mathbb{N}_{0}.\label{eq:lbar-definition}
\end{align}
It then follows that the state evaluated at every $h$ steps is described
by 
\begin{align}
x((i+1)h) & =\Gamma_{\bar{r}(i)}x(ih),\quad i\in\mathbb{N}_{0},\label{eq:lifted-system}
\end{align}
 where 
\begin{align}
\Gamma_{q} & \triangleq A_{q_{h}}A_{q_{h-1}}\cdots A_{q_{1}},\quad q\in\mathcal{M}^{h}.\label{eq:Large-Gamma-Definition}
\end{align}
The dynamical system (\ref{eq:lifted-system}) is a ``lifted'' switched
system with $M^{h}$ number of modes. Each mode of this system is
identified by a sequence of $h$ numbers from $\mathcal{M}$ representing
the modes of the original switched system (\ref{eq:switched-system}). 

Now, let $c_{s}\colon\mathcal{M}^{h}\to\{0,\ldots,h\}$ be defined
by $c_{s}(q)\triangleq\sum_{j=1}^{h}\mathbbm{1}[q_{j}=s],q\in\mathcal{M}^{h},s\in\mathcal{M}.$
With this definition, the number of entries with value $s$ in the
sequence $q\in\mathcal{M}^{h}$ is represented with $c_{s}(q)$. Note
that $c_{s}$ satisfies 
\begin{align}
\sum_{i=0}^{k-1}\sum_{q\in\mathcal{M}^{h}}c_{s}(q)\mathbbm{1}[\bar{r}(i)=q] & =\sum_{i=0}^{kh-1}\mathbbm{1}[r(i)=s],\,\,k\in\mathbb{N},\label{eq:c1-l-l1-relation}
\end{align}
 which establishes a key relation between the mode signal $r(\cdot)$
and the sequence-valued process $\bar{r}(\cdot).$

In Lemma~\ref{Lemma-Limit-Relations} below, we use (\ref{eq:c1-l-l1-relation})
to obtain a relation between $\underline{\rho}_{s},\overline{\rho}_{s}$
in Assumption~\ref{MainAssumption-1} and the long-run average numbers
of the occurrences of all sequences in $\mathcal{M}^{h}$. The long
run average for a sequence $q\in\mathcal{M}^{h}$ is given by 
\begin{align*}
\lim_{k\to\infty}\frac{1}{k}\sum_{i=0}^{k-1}\mathbbm{1}[\bar{r}(i)=q] & ,
\end{align*}
 whenever this limit exists, that is, $\frac{1}{k}\sum_{i=0}^{k-1}\mathbbm{1}[\bar{r}(i)=q]$
converges almost surely to a random variable as $k\to\infty$. 

\begin{lemma} \label{Lemma-Limit-Relations} Suppose $\{r(t)\in\mathcal{M}\}_{t\in\mathbb{N}_{0}}$
satisfies Assumption~\ref{MainAssumption-1} with $\underline{\rho}_{s},\overline{\rho}_{s}\in[0,1]$,
$s\in\mathcal{M}$. If $\lim_{k\to\infty}\frac{1}{k}\sum_{i=0}^{k-1}\mathbbm{1}[\bar{r}(i)=q]$
exists for each $q\in\mathcal{M}^{h}$, then we have 
\begin{align}
 & \sum_{q\in\mathcal{M}^{h}}\frac{c_{s}(q)}{h}\lim_{k\to\infty}\frac{1}{k}\sum_{i=0}^{k-1}\mathbbm{1}[\bar{r}(i)=q]\geq\underline{\rho}_{s},\label{eq:lemma-liminf-term}\\
 & \sum_{q\in\mathcal{M}^{h}}\frac{c_{s}(q)}{h}\lim_{k\to\infty}\frac{1}{k}\sum_{i=0}^{k-1}\mathbbm{1}[\bar{r}(i)=q]\leq\overline{\rho}_{s},\label{eq:lemma-limsup-term}
\end{align}
for $s\in\mathcal{M}$, almost surely. \end{lemma} 

\begin{IEEEproof} We first show (\ref{eq:lemma-limsup-term}). By
(\ref{eq:c1-l-l1-relation}), 
\begin{align}
 & \sum_{q\in\mathcal{M}^{h}}\frac{c_{s}(q)}{h}\lim_{k\to\infty}\frac{1}{k}\sum_{i=0}^{k-1}\mathbbm{1}[\bar{r}(i)=q]\nonumber \\
 & \quad=\lim_{k\to\infty}\frac{1}{kh}\sum_{i=0}^{k-1}\sum_{q\in\mathcal{M}^{h}}c_{s}(q)\mathbbm{1}[\bar{r}(i)=q]\nonumber \\
 & \quad=\lim_{k\to\infty}\frac{1}{kh}\sum_{i=0}^{kh-1}\mathbbm{1}[r(i)=s].\label{eq:limsupdkh}
\end{align}
Here, we have 
\begin{align}
 & \Big\{\frac{1}{\bar{k}h}\sum_{i=0}^{\bar{k}h-1}\mathbbm{1}[r(i)=s]\colon\bar{k}\geq k\Big\}\nonumber \\
 & \quad\subseteq\Big\{\frac{1}{\bar{k}}\sum_{i=0}^{\bar{k}-1}\mathbbm{1}[r(i)=s]\colon\bar{k}\geq k\Big\},\quad k\in\mathbb{N},\label{eq:seteq}
\end{align}
 and hence 
\begin{align}
\sup_{\bar{k}\geq k}\frac{1}{\bar{k}h}\sum_{i=0}^{\bar{k}h-1}\mathbbm{1}[r(i)=s] & \leq\sup_{\bar{k}\geq k}\frac{1}{\bar{k}}\sum_{i=0}^{\bar{k}-1}\mathbbm{1}[r(i)=s],\quad k\in\mathbb{N}.
\end{align}
Therefore, 
\begin{align}
\limsup_{k\to\infty}\frac{1}{kh}\sum_{i=0}^{kh-1}\mathbbm{1}[r(i)=s] & \leq\limsup_{k\to\infty}\frac{1}{k}\sum_{i=0}^{k-1}\mathbbm{1}[r(i)=s].\label{eq:limsupsubseqlessthanlimsup}
\end{align}
As a result, (\ref{eq:lemma-limsup-term}) follows from (\ref{eq:supcond}),
(\ref{eq:limsupdkh}), and (\ref{eq:limsupsubseqlessthanlimsup}). 

The inequality (\ref{eq:lemma-liminf-term}) can be shown using a
similar approach. \end{IEEEproof}

Next, we  employ Lemma~\ref{Lemma-Limit-Relations} to establish
sufficient conditions for almost sure  asymptotic stability. To this
end, first, for a given matrix $N\in\mathbb{R}^{n\times n}$, let
$\|N\|$ denote the induced matrix norm defined by 
\begin{align}
\|N\| & \triangleq\sup_{x\in\mathbb{R}^{n}\setminus\{0\}}\frac{\|Nx\|}{\|x\|},\label{eq:induced-matrix-norm}
\end{align}
 where $\|\cdot\|$ on the right-hand side denotes a vector norm on
$\mathbb{R}^{n}$. In the proof of the next result, we use the submultiplicativity
property of induced matrix norms, i.e., $\|N_{1}N_{2}\|\leq\|N_{1}\|\|N_{2}\|$
for $N_{1},N_{2}\in\mathbb{R}^{n\times n}$ (see Section 5.6 in \cite{hornmatrixanalysis}).

\begin{theorem} \label{Stability-Theorem} Consider the switched
system (\ref{eq:switched-system}). Suppose that the mode signal $\{r(t)\in\{1,\ldots,M\}\}_{t\in\mathbb{N}_{0}}$
satisfies Assumption~\ref{MainAssumption-1} with $\overline{\rho}_{s},\underline{\rho}_{s}\in[0,1]$,
$s\in\mathcal{M}$, and $\lim_{k\to\infty}\frac{1}{k}\sum_{i=0}^{k-1}\mathbbm{1}[\bar{r}(i)=q]$
exists for each $q\in\mathcal{M}^{h}$ for a given $h\in\mathbb{N}$.
If there exist an induced matrix norm $\|\cdot\|$ and a scalar $\varepsilon\in(0,1)$
such that the inequality 
\begin{align}
\sum_{q\in\mathcal{M}^{h}}\gamma_{q}\rho_{q} & <0,\label{eq:maxcondition}
\end{align}
 holds with 
\begin{align}
\gamma_{q} & \triangleq\begin{cases}
\ln\|\Gamma_{q}\|, & \quad\Gamma_{q}\neq0,\\
\ln\varepsilon, & \quad\Gamma_{q}=0,
\end{cases}\quad q\in\mathcal{M}^{h},\label{eq:small-gamma-definition}
\end{align}
for all $\rho_{q}\in[0,1],q\in\mathcal{M}^{h}$, that satisfy 
\begin{align}
\sum_{q\in\mathcal{M}^{h}}\rho_{q} & =1,\label{eq:rhoqineq1}\\
\underline{\rho}_{s}\leq\sum_{q\in\mathcal{M}^{h}}\frac{c_{s}(q)}{h}\rho_{q} & \leq\overline{\rho}_{s},\quad s\in\mathcal{M},\label{eq:rhoqineq2}
\end{align}
 then the zero solution $x(t)\equiv0$ of the dynamical system (\ref{eq:switched-system})
is asymptotically stable almost surely.

\end{theorem} \vskip 3pt

\begin{IEEEproof} First, it follows from (\ref{eq:lifted-system})
that 
\begin{align*}
\|x((k+1)h)\| & =\|\Gamma_{\bar{r}(k)}x(kh)\|\leq\|\Gamma_{\bar{r}(k)}\|\|x(kh)\|,
\end{align*}
 and hence by the submultiplicativity property of the induced matrix
norm $\|\cdot\|$, we have 
\begin{align}
\|x(kh)\| & \leq\eta(k)\|x_{0}\|,\quad k\in\mathbb{N}_{0},\label{eq:x-eta-ineq}
\end{align}
 where $\eta(k)\triangleq\prod_{i=0}^{k-1}\|\Gamma_{\bar{r}(i)}\|$.
Now we define $\mu(k)\triangleq\sum_{i=0}^{k-1}\gamma_{\bar{r}(i)}$,
$k\in\mathbb{N}_{0}$, where $\gamma_{q},q\in\mathcal{M}^{h}$, are
given by (\ref{eq:small-gamma-definition}). It follows from (\ref{eq:small-gamma-definition})
together with the definitions of $\eta(k)$ and $\mu(k)$ that $\eta(k)\leq e^{\mu(k)},k\in\mathbb{N}_{0}$.
Furthermore, since $\gamma_{\bar{r}(i)}=\sum_{q\in\mathcal{M}^{h}}\gamma_{q}\mathbbm{1}[\bar{r}(i)=q]$,
we have 
\begin{align*}
\mu(k) & =\sum_{i=0}^{k-1}\sum_{q\in\mathcal{M}^{h}}\gamma_{q}\mathbbm{1}[\bar{r}(i)=q]=\sum_{q\in\mathcal{M}^{h}}\gamma_{q}\sum_{i=0}^{k-1}\mathbbm{1}[\bar{r}(i)=q],
\end{align*}
for $k\in\mathbb{N},$ and as a result, 
\begin{align}
\lim_{k\to\infty}\frac{1}{k}\mu(k) & =\sum_{q\in\mathcal{M}^{h}}\gamma_{q}\lim_{k\to\infty}\frac{1}{k}\sum_{i=0}^{k-1}\mathbbm{1}[\bar{r}(i)=q],\label{eq:mu-average}
\end{align}
almost surely. Here, note that $\lim_{k\to\infty}\frac{1}{k}\sum_{i=0}^{k-1}\mathbbm{1}[\bar{r}(i)=q]\in[0,1]$.
Furthermore, 
\begin{align}
 & \sum_{q\in\mathcal{M}^{h}}\lim_{k\to\infty}\frac{1}{k}\sum_{i=0}^{k-1}\mathbbm{1}[\bar{r}(i)=q]\nonumber \\
 & \quad=\lim_{k\to\infty}\frac{1}{k}\sum_{i=0}^{k-1}\sum_{q\in\mathcal{M}^{h}}\mathbbm{1}[\bar{r}(i)=q]=\lim_{k\to\infty}\frac{1}{k}\sum_{i=0}^{k-1}1=1.\label{eq:sum1}
\end{align}
Let $\rho_{q}^{*}\triangleq\lim_{k\to\infty}\frac{1}{k}\sum_{i=0}^{k-1}\mathbbm{1}[\bar{r}(i)=q]$,
$q\in\mathcal{M}^{h}$, and 
\begin{align*}
\vartheta\triangleq\max\Big\{\sum_{q\in\mathcal{M}^{h}}\gamma_{q}\rho_{q}\colon\rho_{q}\in[0,1],q\in\mathcal{M}^{h},\,\text{s.t.}\,\eqref{eq:rhoqineq1},\eqref{eq:rhoqineq2}\Big\}.
\end{align*}
 Furthermore, let $E,F\subset\mathcal{F}$ be the events defined by
$ $
\begin{align*}
E & \triangleq\{\omega\in\Omega\colon\sum_{q\in\mathcal{M}^{h}}\rho_{q}^{*}=1,\\
 & \qquad\underline{\rho}_{s}\leq\sum_{q\in\mathcal{M}^{h}}\frac{c_{s}(q)}{h}\rho_{q}^{*}\leq\overline{\rho}_{s},\,s\in\mathcal{M}\}.\\
F & \triangleq\{\omega\in\Omega\colon\sum_{q\in\mathcal{M}^{h}}\gamma_{q}\rho_{q}^{*}\leq\vartheta\}.
\end{align*}
We first show that $\mathbb{P}[E]=1$. Observe that (\ref{eq:sum1})
implies (\ref{eq:rhoqineq1}) with $\rho_{q}$ replaced by $\rho_{q}^{*}$.
Moreover, Lemma~\ref{Lemma-Limit-Relations} implies that (\ref{eq:rhoqineq2})
with $\rho_{q}$ replaced by $\rho_{q}^{*}$ holds almost surely.
Hence, we have $\mathbb{P}[E]=1$. Now, since $E\subseteq F$, we
also have $\mathbb{P}[F]=1$. Furthermore, (\ref{eq:maxcondition})
implies that $\vartheta<0$. As a result, by noting that $\mathbb{P}[F]=1$,
we obtain $\sum_{q\in\mathcal{M}^{h}}\gamma_{q}\rho_{q}^{*}\leq\vartheta<0$,
almost surely. We use this fact together with (\ref{eq:mu-average})
to obtain 
\begin{align}
\lim_{k\to\infty}\frac{1}{k}\mu(k) & =\sum_{q\in\mathcal{M}^{h}}\gamma_{q}\rho_{q}^{*}\leq\vartheta<0,\label{eq:muklimit}
\end{align}
 almost surely. Here, (\ref{eq:muklimit}) implies $\lim_{k\to\infty}\mu(k)=-\infty$,
almost surely. As a result, by noting that $\eta(k)\leq e^{\mu(k)}$,
we obtain $\mathbb{P}[\lim_{k\to\infty}\eta(k)=0]=1$. Thus, for any
$\epsilon>0$, $\lim_{j\to\infty}\mathbb{P}[\sup_{k\geq j}\eta(k)>\epsilon]=0$
(see Proposition~5.6 of \cite{karrprobabilitybook}). Therefore,
for any $\epsilon>0$ and $\bar{p}>0$, there exists a positive integer
$N(\epsilon,\bar{p})$ such that 
\begin{align}
\mathbb{P}[\sup_{k\geq j}\eta(k) & >\epsilon]<\bar{p},\quad j\geq N(\epsilon,\bar{p}).\label{eq:supetainequalit}
\end{align}

In what follows, we show almost sure stability of the switched system
by using (\ref{eq:x-eta-ineq}) and (\ref{eq:supetainequalit}). First,
we define 
\begin{align}
\phi & \triangleq\max\{1,\max_{s\in\mathcal{M}}\|A_{s}\|\}\label{eq:phidef}
\end{align}
 and $\mathcal{T}_{k}\triangleq\{kh,\ldots,(k+1)h-1\}$, $k\in\mathbb{N}_{0}$.
Using these definitions, we obtain 
\begin{align}
\|x(t+1)\| & =\|A_{r(t)}x(t)\|\leq\|A_{r(t)}\|\|x(t)\|\nonumber \\
 & \leq\phi\|x(t)\|,\quad t\in\mathcal{T}_{k}.\label{eq:xphiineq}
\end{align}
 It then follows from (\ref{eq:xphiineq}) that $\|x(t)\|\leq\phi^{t-kh}\|x(kh)\|$,
$t\in\mathcal{T}_{k}$. Since $\mathcal{T}_{k}$ has $h$ time instants
and $\phi\geq1$, we have $\phi^{t-kh}\leq\phi^{h-1}\leq\phi^{h}$
and hence $\|x(t)\|\leq\phi^{h}\|x(kh)\|$ for all $t\in\mathcal{T}_{k}$.
Consequently, 
\begin{align}
\max_{t\in\mathcal{T}_{k}}\|x(t)\| & \leq\phi^{h}\|x(kh)\|,\quad k\in\mathbb{N}_{0}.\label{eq:max-x-ineq}
\end{align}
Now by (\ref{eq:x-eta-ineq}) and (\ref{eq:max-x-ineq}), 
\begin{align*}
\eta(k) & \geq\|x(kh)\|\|x_{0}\|^{-1}\geq\max_{t\in\mathcal{T}_{k}}\|x(t)\|\phi^{-h}\|x_{0}\|^{-1},\,k\in\mathbb{N}_{0}.
\end{align*}
Then it follows from (\ref{eq:supetainequalit}) that for all $\epsilon>0$
and $\bar{p}>0$, 
\begin{align*}
 & \mathbb{P}[\sup_{k\geq j}\max_{t\in\mathcal{T}_{k}}\|x(t)\|>\epsilon\phi^{h}\|x_{0}\|]\\
 & \quad=\mathbb{P}[\sup_{k\geq j}\max_{t\in\mathcal{T}_{k}}\|x(t)\|\phi^{-h}\|x_{0}\|^{-1}>\epsilon]\\
 & \quad\leq\mathbb{P}[\sup_{k\geq j}\eta(k)>\epsilon]<\bar{p},\quad j\geq N(\epsilon,\bar{p}).
\end{align*}
Now let $\delta_{1}\triangleq\phi^{-h}$. Notice that if $\|x_{0}\|\leq\delta_{1}$,
then $\phi^{h}\|x_{0}\|\leq1$, and therefore, for all $j\geq N(\epsilon,\bar{p})$,
we have 
\begin{align}
 & \mathbb{P}[\sup_{k\geq j}\max_{t\in\mathcal{T}_{k}}\|x(t)\|>\epsilon]\nonumber \\
 & \quad\leq\mathbb{P}[\sup_{k\geq j}\max_{t\in\mathcal{T}_{k}}\|x(t)\|>\epsilon\phi^{h}\|x_{0}\|]<\bar{p}.\label{eq:epsilon-result-part1}
\end{align}
Furthermore, observe that for all $k\in\{0,1,\ldots,N(\epsilon,\bar{p})-1\}$,
we have $\|x(kh)\|\leq\phi^{k}\|x_{0}\|\leq\phi^{N(\epsilon,\bar{p})-1}\|x_{0}\|$.
Hence, as a result of (\ref{eq:max-x-ineq}), 
\begin{align}
\max_{t\in\mathcal{T}_{k}}\|x(t)\| & \leq\phi^{h}\|x(kh)\|\leq\phi^{h+N(\epsilon,\bar{p})-1}\|x_{0}\|,\label{eq:max-n-ineq}
\end{align}
for all $k\in\{0,1,\ldots,N(\epsilon,\bar{p})-1\}$. Let $\delta_{2}(\epsilon,\bar{p})\triangleq\epsilon\phi^{1-h-N(\epsilon,\bar{p})}$.
Now, if $\|x_{0}\|\leq\delta_{2}(\epsilon,\bar{p})$, then by (\ref{eq:max-n-ineq}),
$\max_{t\in\mathcal{T}_{k}}\|x(t)\|\leq\epsilon$, $k\in\{0,1,\ldots,N(\epsilon,\bar{p})-1\}$.
Thus, if $\|x_{0}\|\leq\delta_{2}(\epsilon,\bar{p})$, then 
\begin{eqnarray}
\mathbb{P}[\max_{k\in\{0,1,\ldots,N(\epsilon,\bar{p})\}}\max_{t\in\mathcal{T}_{k}}\|x(t)\|>\epsilon] & = & 0.\label{eq:epsilon-result-part2}
\end{eqnarray}
 Due to (\ref{eq:epsilon-result-part1}) and (\ref{eq:epsilon-result-part2}),
for all $\epsilon>0$, $\bar{p}>0$, we have 
\begin{align}
 & \mathbb{P}[\sup_{t\in\mathbb{N}_{0}}\|x(t)\|>\epsilon]=\mathbb{P}[\sup_{k\in\mathbb{N}_{0}}\max_{t\in\mathcal{T}_{k}}\|x(t)\|>\epsilon]\nonumber \\
 & \quad=\mathbb{P}[\{\max_{k\in\{0,\ldots,N(\epsilon,\bar{p})-1\}}\max_{t\in\mathcal{T}_{k}}\|x(t)\|>\epsilon\}\nonumber \\
 & \quad\quad\quad\cup\,\{\sup_{k\geq N(\epsilon,\bar{p})}\max_{t\in\mathcal{T}_{k}}\|x(t)\|>\epsilon\}]\nonumber \\
 & \quad\leq\mathbb{P}[\max_{k\in\{0,\ldots,N(\epsilon,\bar{p})-1\}}\max_{t\in\mathcal{T}_{k}}\|x(t)\|>\epsilon]\nonumber \\
 & \quad\quad+\mathbb{P}[\sup_{k\geq N(\epsilon,\bar{p})}\max_{t\in\mathcal{T}_{k}}\|x(t)\|>\epsilon]<\bar{p},\label{eq:almost-sure-stability-one-step-before}
\end{align}
 whenever $\|x_{0}\|<\min(\delta_{1},\delta_{2}(\epsilon,\bar{p}))$. 

By Corollary 5.4.5 of \cite{hornmatrixanalysis}, there exist $c_{1},c_{2}>0$
such that 
\begin{align}
c_{1}\|x\|\leq\|x\|_{2} & \leq c_{2}\|x\|,\quad x\in\mathbb{R}^{n}.\label{eq:norm-equality}
\end{align}
By (\ref{eq:almost-sure-stability-one-step-before}) and (\ref{eq:norm-equality}),
we obtain that for all $\epsilon>0$, $\bar{p}>0$, 
\begin{align*}
\mathbb{P}[\sup_{t\in\mathbb{N}_{0}}\|x(t)\|_{2}>\epsilon] & \leq\mathbb{P}[\sup_{t\in\mathbb{N}_{0}}\|x(t)\|>\frac{\epsilon}{c_{2}}]<\bar{p},
\end{align*}
 whenever $\|x_{0}\|<\min(\delta_{1},\delta_{2}(\frac{\epsilon}{c_{2}},\bar{p}))$.
Now, since $\|x_{0}\|\leq\frac{\|x_{0}\|_{2}}{c_{1}}$, we have that
for all $\epsilon>0$, $\bar{p}>0$, the inequality (\ref{eq:definition-stability})
holds whenever $\|x_{0}\|_{2}<\delta(\epsilon,\bar{p})\triangleq c_{1}\min(\delta_{1},\delta_{2}(\frac{\epsilon}{c_{2}},\bar{p}))$,
which implies almost sure stability. 

Next, we show (\ref{eq:definition-convergence}) to establish almost
sure \emph{asymptotic} stability of the zero solution. In this regard,
first notice that $\mathbb{P}[\lim_{k\to\infty}\eta(k)=0]=1$. By
using (\ref{eq:x-eta-ineq}), we obtain $\mathbb{P}[\lim_{k\to\infty}\|x(kh)\|=0]=1$,
which implies $\mathbb{P}[\lim_{t\to\infty}\|x(t)\|=0]=1$. Now as
a consequence of (\ref{eq:norm-equality}), we have (\ref{eq:definition-convergence}).
Hence the zero solution of the switched system (\ref{eq:switched-system})
is asymptotically stable almost surely. \end{IEEEproof}

Theorem~\ref{Stability-Theorem} provides an almost sure asymptotic
stability condition for the switched system (\ref{eq:switched-system}).
This result indicates that the stability can be assessed by checking
the inequality (\ref{eq:maxcondition}) for all scalars $\rho_{q}\in[0,1],q\in\mathcal{M}^{h}$,
such that (\ref{eq:rhoqineq1}), (\ref{eq:rhoqineq2}) hold. In (\ref{eq:maxcondition}),
the scalar $\gamma_{q}\in\mathbb{R}$ represents the effect of mode
sequence $q$. Specifically, for mode sequences with $\|\Gamma_{q}\|<1$,
we have $\gamma_{q}<0$. A negative value for $\gamma_{q}$ implies
that the norm of the system's state gets smaller after $h$ time steps,
if the mode within those $h$ time steps follows the sequence $q$.
Note that for the case $\Gamma_{q}=0$, $\varepsilon\in(0,1)$ in
(\ref{eq:small-gamma-definition}) ensures that $\gamma_{q}$ is well
defined and negative. In practice, $\varepsilon$ can be selected
as a very small positive number. On the other hand, for mode sequences
with $\|\Gamma_{q}\|>1$, we have $\gamma_{q}>0$. A positive value
for $\gamma_{q}$ indicates that the mode sequence $q$ may cause
the norm of the system's state to increase. Hence, the term $\sum_{q\in\mathcal{M}^{h}}\gamma_{q}\rho_{q}$
in (\ref{eq:maxcondition}) with $\rho_{q}=\lim_{k\to\infty}\frac{1}{k}\sum_{i=0}^{k-1}\mathbbm{1}[\bar{r}(i)=q]$
would correspond to the average of the effects of all $h$-length
sequences in $\mathcal{M}^{h}$. However, this average cannot be computed
directly, since in this paper, we consider the case where the specific
values of $\lim_{k\to\infty}\frac{1}{k}\sum_{i=0}^{k-1}\mathbbm{1}[\bar{r}(i)=q]$
are not available. 

On the other hand, we show by using Lemma~\ref{Lemma-Limit-Relations}
that if the long-run average activity of modes is known to be bounded
as in (\ref{eq:infcond}) and (\ref{eq:supcond}), then $\rho_{q}=\lim_{k\to\infty}\frac{1}{k}\sum_{i=0}^{k-1}\mathbbm{1}[\bar{r}(i)=q]$
would satisfy (\ref{eq:rhoqineq1}), (\ref{eq:rhoqineq2}). Hence,
for stability analysis, one can check the sign of $\sum_{q\in\mathcal{M}^{h}}\gamma_{q}\rho_{q}$
in (\ref{eq:maxcondition}) for all $\rho_{q}\in[0,1],q\in\mathcal{M}^{h},$
that satisfy (\ref{eq:rhoqineq1}), (\ref{eq:rhoqineq2}). This is
equivalent to checking the stability for all possible mode sequence
scenarios that satisfy (\ref{eq:infcond}) and (\ref{eq:supcond}),
since different values of the limits $\rho_{q}=\lim_{k\to\infty}\frac{1}{k}\sum_{i=0}^{k-1}\mathbbm{1}[\bar{r}(i)=q]$,
$q\in\mathcal{M}^{h}$, represent different scenarios. We will show
in Section~\ref{sec:Linear-Programming-Methods} that rather than
checking the condition in (\ref{eq:maxcondition}) for all possible
scenarios, we can utilize linear programming methods to identify the
worst scenario in terms of stability, and check the condition only
for that scenario. 

Note that in Theorem~\ref{Stability-Theorem} we require the existence
of $\lim_{k\to\infty}\frac{1}{k}\sum_{i=0}^{k-1}\mathbbm{1}[\bar{r}(i)=q]$
for all $q\in\mathcal{M}^{h}$, even though the particular values
of these limits are not needed for stability analysis. The following
result identifies a class of mode signals $\{r(t)\in\{0,1\}\}_{t\in\mathbb{N}_{0}}$
for which these limits exist. Using this result, we will show that
our analysis technique is applicable in a variety of scenarios.

\begin{proposition} \label{Proposition-l-g} Let $\{g(t)\in\mathcal{S}\}_{t\in\mathbb{N}_{0}}$
with $g(0)=g_{0}\in\mathcal{S}$ be a finite-state irreducible Markov
chain. Assume $\{r(t)\in\mathcal{M}\}_{t\in\mathbb{N}_{0}}$ is given
by 
\begin{align}
r(t) & \triangleq\begin{cases}
1,\quad & g(t)\in\mathcal{S}_{1},\\
\quad\quad\vdots\\
M,\quad & g(t)\in\mathcal{S}_{M},
\end{cases}\quad t\in\mathbb{N}_{0},\label{eq:l-g-definition}
\end{align}
 where $\mathcal{S}_{1},\mathcal{S}_{2},\ldots,\mathcal{S}_{M}$ form
a partition of the set $\mathcal{S}$, i.e., $\cup_{i=1}^{M}\mathcal{S}_{i}=\mathcal{S}$
and $\mathcal{S}_{i}\cap\mathcal{S}_{j}=\emptyset$, $i\neq j$. Then
for all $h\in\mathbb{N}$, the limits $\lim_{k\to\infty}\frac{1}{k}\sum_{i=0}^{k-1}\mathbbm{1}[\bar{l}(i)=q]$,
$q\in\mathcal{M}^{h}$, exist. 

\end{proposition}

 The proof of this result is composed of several key steps.
The first step is based on the observation  that $\bar{r}(\cdot)$ is generated from sequences of values that
the process $r(\cdot)$ takes between every $h$ time steps. By exploiting
this observation, we construct a new process $\bar{g}(\cdot)$ representing
the sequences of values that $g(\cdot)$ takes between every $d$
steps, where $d$ is a carefully chosen period length that is an integer
multiple of $h$. In the second step, we establish the relation between
the processes $\bar{r}(\cdot)$ and $\bar{g}(\cdot)$ by using (\ref{eq:l-g-definition}).
Then, in the final step, we show that $\lim_{k\to\infty}\frac{1}{k}\sum_{i=0}^{k-1}\mathbbm{1}[\bar{r}(i)=q]$
can be obtained by utilizing invariant distributions of $\bar{g}(\cdot)$. 

\begin{inarxiv}

\begin{IEEEproof} In the proof, we use the notion of \emph{period}
for Markov chains \cite{rosenthal2006first}. Specifically, the period
$\tau_{\sigma}\in\mathbb{N}$ of a state $\sigma\in\mathcal{S}$ is
defined by 
\begin{align*}
\tau_{\sigma}\triangleq\gcd\{t\in\mathbb{N}\colon\mathbb{P}[g(t)=\sigma|g(0)=\sigma]>0\}, & \quad\sigma\in\mathcal{S},
\end{align*}
 where $\mathrm{gcd}(T)$ denotes the greatest common denominator
of the elements of the set $T$. By this definition, the random time
intervals between revisits to state $\sigma$ are guaranteed to be
integer multiples of $\tau_{\sigma}$. Since $\{g(t)\in\mathcal{S}\}_{t\in\mathbb{N}_{0}}$
is an irreducible finite-state Markov chain, it follows from Corollary
8.3.7 of \cite{rosenthal2006first} that the period is the same for
all states. We use $\tau\in\mathbb{N}$ to denote this period, i.e.,
$\tau=\tau_{1}=\tau_{2}=\cdots=\tau_{|\mathcal{S}|}$, where $|\mathcal{S}|$
denotes the number of elements in the set $\mathcal{S}$. Now let
$d\triangleq\tau h$. 

Next, we define a sequence-valued process to characterize the evolution
of $g(\cdot)$ in every $d$ steps. To this end, first, for each $\sigma\in\mathcal{S}$,
let $\mathcal{I}_{\sigma,k}\triangleq\{s\in\mathcal{S}\colon\mathbb{P}[g(kd)=s|g(0)=\sigma]>0\}$,
$k\in\{1,\ldots,|\mathcal{S}|\}$, and $\mathcal{I}_{\sigma}\triangleq\cup_{k=1}^{|\mathcal{S}|}\mathcal{I}_{\sigma,k}$.
The set $\mathcal{I}_{\sigma}\subset\mathcal{S}$ denotes the states
that can be reached from the state $\sigma$ in steps that are integer
multiples of $d$. In addition, for each $\sigma\in\mathcal{S}$,
let 
\begin{align*}
\bar{\mathcal{S}}_{\sigma} & \triangleq\{(\bar{s}_{1},\bar{s}_{2},\ldots,\bar{s}_{d})\colon\bar{s}_{j}\in\mathcal{S},j\in\{1,\ldots,d\},\bar{s}_{1}\in\mathcal{I}_{\sigma},\\
 & \quad\mathbb{P}[g(1)=\bar{s}_{2},\ldots,g(d-1)=\bar{s}_{d}|g(0)=\bar{s}_{1}]>0\}.
\end{align*}
 Now we define the sequence-valued process $\{\bar{g}(i)\}_{i\in\mathbb{N}_{0}}$
by 
\begin{align}
\bar{g}(i) & \triangleq(g(id),g(id+1),\ldots,g((i+1)d-1)).\label{eq:gbar-definition-1}
\end{align}
Notice that $\bar{g}(i)\in\bar{\mathcal{S}}_{g_{0}}$, $i\in\mathbb{N}_{0}$. 

Our next goal is to show that the sequence-valued Markov chain $\{\bar{g}(i)\in\bar{\mathcal{S}}_{g_{0}}\}_{i\in\mathbb{N}_{0}}$
is an irreducible Markov chain. Specifically, we prove that for every
$\bar{\sigma},\bar{s}\in\bar{\mathcal{S}}_{g_{0}}$, there exists
$\bar{k}\in\mathbb{N}$ such that 
\begin{align}
 & \mathbb{P}[\bar{g}(i+\bar{k})=\bar{s}\,|\,\bar{g}(i)=\bar{\sigma}]>0.\label{eq:rbar-sbar}
\end{align}
To this end, first note that $\bar{g}_{1}(i)\in\mathcal{I}_{g_{0}}$,
$i\in\mathbb{N}_{0}$, that is, the first elements of the sequence-values
that $\bar{g}(\cdot)$ takes are elements of the set $\mathcal{I}_{g_{0}}$.
It follows from the definition of $\mathcal{I}_{g_{0}}$ that for
all $\sigma,s\in\mathcal{I}_{g_{0}}$, 
\begin{align*}
\{k\in\mathbb{N}\colon\mathbb{P}[g(kd)=s|g(0)=\sigma]>0\} & \neq\emptyset.
\end{align*}
Now, define $k\colon\mathcal{I}_{g_{0}}\times\mathcal{I}_{g_{0}}\to\mathbb{N}$
by 
\begin{align}
k(\sigma,s) & \triangleq\min\{k\in\mathbb{N}\colon\mathbb{P}[g(kd)=s|g(0)=\sigma]>0\}.\label{eq:kdefinition}
\end{align}
Moreover, note that for any given $\bar{\sigma},\bar{s}\in\bar{\mathcal{S}}_{g_{0}}$,
we can always pick a state $c\in\mathcal{I}_{g_{0}}$ and let $\bar{k}\triangleq k(c,\bar{s}_{1})+1$
so that 
\begin{align*}
 & \mathbb{P}[g(1)=c|g(0)=\bar{\sigma}_{d}]>0.
\end{align*}
By (\ref{eq:kdefinition}), we have $\mathbb{P}[g(k(c,\bar{s}_{1})d)=\bar{s}_{1}|g(0)=c]>0$,
and consequently 
\begin{align*}
 & \mathbb{P}[\bar{g}(i+\bar{k})=\bar{s}|\bar{g}(i)=\bar{\sigma}]\\
 & \quad=\mathbb{P}[\bar{g}(i+\bar{k})=\bar{s}|g((i+1)d-1)=\bar{\sigma}_{d}]\\
 & \quad=\mathbb{P}[g((i+\bar{k})d+1)=\bar{s}_{2},\ldots,\quad\\
 & \quad\quad\quad g((i+\bar{k}+1)d-1)=\bar{s}_{d}|g((i+\bar{k})d)=\bar{s}_{1}]\\
 & \quad\quad\cdot\mathbb{P}[g((i+\bar{k})d)=\bar{s}_{1}|g((i+1)d-1)=\bar{\sigma}_{d}]\\
 & \quad\geq\mathbb{P}[g((i+\bar{k})d+1)=\bar{s}_{2},\ldots,\\
 & \quad\quad\quad g((i+\bar{k}+1)d-1)=\bar{s}_{d}|g((i+\bar{k})d)=\bar{s}_{1}]\\
 & \quad\quad\cdot\mathbb{P}[g((i+\bar{k})d)=\bar{s}_{1}|g((i+1)d)=c]\\
 & \quad\quad\cdot\mathbb{P}[g((i+1)d)=c|g((i+1)d-1)=\bar{\sigma}_{d}]>0.
\end{align*}
 Thus, the sequence-valued Markov chain $\{\bar{g}(i)\in\bar{\mathcal{S}}_{g_{0}}\}_{i\in\mathbb{N}_{0}}$
is irreducible. Now, define the function $\alpha:\mathcal{\bar{S}}_{g_{0}}\times\mathcal{M}^{h}\to\mathbb{N}_{0}$
by 
\begin{align}
\alpha(\bar{s},q) & \triangleq\sum_{j=0}^{\tau-1}\mathbbm{1}[\bar{s}_{jh+1}\in\mathcal{S}_{q_{1}},\ldots,\bar{s}_{jh+h}\in\mathcal{S}_{q_{h}}],\label{eq:alpha-definition}
\end{align}
 for $\bar{s}\in\mathcal{\bar{S}}_{g_{0}},q\in\mathcal{M}^{h}$. Note
that $\alpha(\bar{s},q)\in\mathbb{N}_{0}$ is the number of times
the sequence $q$ appears in the process $r(\cdot)$, when the process
$g(\cdot)$ takes the values $\bar{s}_{1},\bar{s}_{2},\ldots,\bar{s}_{d}$.
This number is computed by dividing the $d$-length sequence $\bar{s}$
into $\tau$ number of $h$-length sequences and counting the number
of $h$-length sequences whose elements are from sets $\mathcal{S}_{q_{1}},\mathcal{S}_{q_{2}},\ldots,\mathcal{S}_{q_{h}}$. 

By using (\ref{eq:alpha-definition}), we get 
\begin{eqnarray}
\frac{1}{\tau k}\sum_{i=0}^{\tau k-1}\mathbbm{1}[\bar{r}(i)=q]=\frac{1}{\tau k}\sum_{i=0}^{k-1}\sum_{\bar{s}\in\mathcal{\bar{S}}_{g_{0}}}\mathbbm{1}[\bar{g}(i)=\bar{s}]\alpha(\bar{s},q). &  & \,\,\label{eq:lim-dk}
\end{eqnarray}
 Let $\bar{\pi}_{g_{0},\bar{s}}\in[0,1]$, $\bar{s}\in\bar{\mathcal{S}}_{g_{0}}$,
denote the invariant distribution associated with $\{\bar{g}(i)\in\bar{\mathcal{S}}_{g_{0}}\}_{i\in\mathbb{N}_{0}}$.
It then follows from ergodic theorem for finite-state Markov chains
(see Theorem 1.10.2 of \cite{norris2009}) that 
\begin{align*}
\lim_{k\to\infty}\frac{1}{k}\sum_{i=0}^{k-1}\sum_{\bar{s}\in\mathcal{\bar{S}}_{g_{0}}}\mathbbm{1}[\bar{g}(i) & =\bar{s}]\alpha(\bar{s},q)=\sum_{\bar{s}\in\mathcal{\bar{S}}_{g_{0}}}\bar{\pi}_{g_{0},\bar{s}}\alpha(\bar{s},q).
\end{align*}
 Hence, as a consequence of (\ref{eq:lim-dk}), $\lim_{k\to\infty}\frac{1}{\tau k}\sum_{i=0}^{\tau k-1}\mathbbm{1}[\bar{r}(i)=q]$
exists and is given by 
\begin{align}
\lim_{k\to\infty}\frac{1}{\tau k}\sum_{i=0}^{\tau k-1}\mathbbm{1}[\bar{r}(i)=q] & =\frac{1}{\tau}\sum_{\bar{s}\in\mathcal{\bar{S}}_{g_{0}}}\bar{\pi}_{g_{0},\bar{s}}\alpha(\bar{s},q).\label{eq:dalpha}
\end{align}

Our final goal is to show 
\begin{align*}
\lim_{k\to\infty}\frac{1}{k}\sum_{i=0}^{k-1}\mathbbm{1}[\bar{r}(i)=q] & =\lim_{k\to\infty}\frac{1}{\tau k}\sum_{i=0}^{\tau k-1}\mathbbm{1}[\bar{r}(i)=q].
\end{align*}
To this end, first let $\theta(k)\triangleq\lfloor\frac{k}{\tau}\rfloor$
and observe that 
\begin{align}
 & \frac{1}{k}\sum_{i=0}^{k-1}\mathbbm{1}[\bar{r}(i)=q]=\frac{1}{k}\sum_{i=0}^{\tau\theta(k)-1}\mathbbm{1}[\bar{r}(i)=q]\nonumber \\
 & \quad+\frac{1}{k}\sum_{i=\tau\theta(k)}^{k-1}\mathbbm{1}[\bar{r}(i)=q],\quad k\in\mathbb{N}_{0}.\label{eq:sum-before-limits}
\end{align}
Since $\mathbbm{1}[\bar{r}(i)=q]\in\{0,1\}$, we have $0\leq\frac{1}{k}\sum_{i=\tau\theta(k)}^{k-1}\mathbbm{1}[\bar{r}(i)=q]\leq\frac{\tau}{k},$
and hence, $0\leq\lim_{k\to\infty}\frac{1}{k}\sum_{i=\tau\theta(k)}^{k-1}\mathbbm{1}[\bar{r}(i)=q]\leq\lim_{k\to\infty}\frac{\tau}{k}=0$.
As a result, 
\begin{align}
\lim_{k\to\infty}\frac{1}{k}\sum_{i=\tau\theta(k)}^{k-1}\mathbbm{1}[\bar{r}(i) & =q]=0.\label{eq:zero-limit}
\end{align}
 Furthermore, we have 
\begin{align}
 & \lim_{k\to\infty}\frac{1}{k}\sum_{i=0}^{\tau\theta(k)-1}\mathbbm{1}[\bar{r}(i)=q]\nonumber \\
 & \quad=\lim_{k\to\infty}\frac{1}{k}\frac{\tau\theta(k)}{\tau\theta(k)}\sum_{i=0}^{\tau\theta(k)-1}\mathbbm{1}[\bar{r}(i)=q]\nonumber \\
 & \quad=\lim_{k\to\infty}\frac{\tau\theta(k)}{k}\lim_{k\to\infty}\frac{1}{\tau\theta(k)}\sum_{i=0}^{\tau\theta(k)-1}\mathbbm{1}[\bar{r}(i)=q].\label{eq:dtheta}
\end{align}
 Since $\lim_{k\to\infty}\frac{\tau\theta(k)}{k}=1$, by (\ref{eq:dalpha})
and (\ref{eq:dtheta}), 
\begin{align}
\lim_{k\to\infty}\frac{1}{k}\sum_{i=0}^{\tau\theta(k)-1}\mathbbm{1}[\bar{r}(i)=q] & =\frac{1}{\tau}\sum_{\bar{s}\in\mathcal{\bar{S}}_{g_{0}}}\bar{\pi}_{g_{0},\bar{s}}\alpha(\bar{s},q).\label{eq:dtheta-limit}
\end{align}
 It then follows from (\ref{eq:sum-before-limits}), (\ref{eq:zero-limit}),
and (\ref{eq:dtheta-limit}) that $\lim_{k\to\infty}\frac{1}{k}\sum_{i=0}^{k-1}\mathbbm{1}[\bar{r}(i)=q]$
exists and is given by 
\begin{align*}
\lim_{k\to\infty}\frac{1}{k}\sum_{i=0}^{k-1}\mathbbm{1}[\bar{r}(i)=q] & =\frac{1}{\tau}\sum_{\bar{s}\in\mathcal{\bar{S}}_{g_{0}}}\bar{\pi}_{g_{0},\bar{s}}\alpha(\bar{s},q).
\end{align*}
\end{IEEEproof}

\end{inarxiv}

\begin{remark} In Proposition~\ref{Proposition-l-g},  we provide
a characterization of the mode signal $\{r(t)\in\{1,\ldots,M\}\}_{t\in\mathbb{N}_{0}}$
through an irreducible Markov chain $\{g(t)\in\mathcal{S}\}_{t\in\mathbb{N}_{0}}$.
The set $\mathcal{S}$ of the possible values of $g(\cdot)$ is the
union of disjoint sets $\mathcal{S}_{1},\mathcal{S}_{2},\ldots,\mathcal{S}_{M}$.
By the definition in (\ref{eq:l-g-definition}), the mode signal takes
the value $s$, when $g(t)\in\mathcal{S}_{s}$. Observe that this
mode signal follows the hidden Markov model \cite{vidyasagar2014hidden},
and even though $\{g(t)\in\mathcal{S}\}_{t\in\mathbb{N}_{0}}$ is
a Markov chain, $\{r(t)\in\mathcal{M}\}_{t\in\mathbb{N}_{0}}$ may
fail to be so as we will show in two examples shortly below. To see
this, we first express the switched system (\ref{eq:switched-system})
equivalently by the alternative system 
\begin{align}
x(t+1) & =\bar{A}_{g(t)}x(t),\label{eq:mjs}
\end{align}
 with $\bar{A}_{j}\triangleq A_{i}$, $j\in\mathcal{S}_{i}$. Here,
(\ref{eq:mjs}) is a switched system where the mode signal $\{g(t)\in\mathcal{S}\}_{t\in\mathbb{N}_{0}}$
is a Markov chain. This switched system is also called a Markov jump
system \cite{costa2004discrete}. We would like to highlight that
the stability results on Markov jump systems are not applicable here.
Different from the ordinary Markov jump systems, neither transition
probabilities nor stationary distributions of the process $g(\cdot)$
are available for analysis. In fact even the size of the set $\mathcal{S}$
may not be known. In particular, when we consider the application
to networked control under jamming attacks (see Sections~\ref{sec:Application-to-Networked}
and \ref{sec:Numerical-Example}), the process $\{g(t)\in\mathcal{S}\}_{t\in\mathbb{N}_{0}}$
may characterize a jamming attacker's strategy and its properties
are not available for analysis. 

\begin{inarxiv}Furthermore, we remark that (\ref{eq:l-g-definition})
is only one of the possible characterizations of the mode signal for
our analysis to be applicable. The mode signal may also be generated
in other ways following different random or deterministic characterizations.
Our stability analysis in Theorem~\ref{Stability-Theorem} relies
on the bounds on the average number of times each mode is active in
the long run (see Assumption~\ref{MainAssumption-1}) instead of
the properties of particular mode signal characterizations. \end{inarxiv}
\end{remark}

\begin{figure}
\begin{center}\includegraphics[width=0.8\columnwidth]{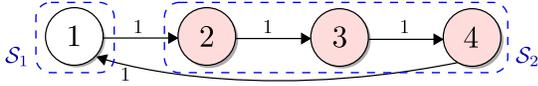}\end{center}
\vskip -5pt

\caption{Transition diagram for Markov chain $\{g(t)\in\mathcal{S}=\{1,2,3,4\}\}_{t\in\mathbb{N}_{0}}$
with initial condition $g(0)=1$ and partitions $\mathcal{S}_{1}=\{1\}$,
$\mathcal{S}_{2}=\{2,3,4\}$, characterizing a $4$-periodic mode
sequence $1,2,2,2,1,2,2,2,\ldots$.}
 \label{Flo:Periodic-Chain} \vskip -12pt
\end{figure}

The characterization through (\ref{eq:l-g-definition}) is general
enough to model various types of mode signals. We present two examples
in this regard. 

\emph{Example 1:} Periodic mode switchings can be described with an
irreducible and periodic Markov chain $\{g(t)\in\mathcal{S}\}_{t\in\mathbb{N}_{0}}$.
Consider for example a switched system with $2$ modes. The mode sequence
is assumed to repeat itself in every $4$ time steps. Specifically,
in every $4$ time steps, mode $1$ is active for $1$ time step then
mode $2$ becomes active for the next $3$ time steps. This switching
scenario can be characterized by setting $\mathcal{S}_{1}\triangleq\{1\}$,
$\mathcal{S}_{2}\triangleq\{2,3,4\}$, and $\{g(t)\in\mathcal{S}\}_{t\in\mathbb{N}_{0}}$
as a Markov chain with transition probabilities shown on the edges
of the transition graph in Fig.~\ref{Flo:Periodic-Chain}. Here,
$g(\cdot)$ repeatedly takes the values $1,2,3,4,1,2,3,4,\ldots$.
As a result, by the definition in (\ref{eq:l-g-definition}), the
mode signal $r(\cdot)$ takes the values $1,2,2,2,1,2,2,2,\ldots$
indicating the periodic change in the mode. 

In this example, the switched system can be used for modeling a networked
control system under periodic attacks. In particular, the first mode
corresponds to a successful packet exchange between the plant and
the controller, and the second mode represents the dynamics when there
is a transmission failure due to attacks. The characterization of
the mode signal through the setup in Fig.~\ref{Flo:Periodic-Chain}
represents an example of the discrete-time version of the periodic
attacks discussed in \cite{HSF-SM:ct-13}. Here the attacker periodically
repeats sleeping for $1$ time step and emitting a jamming signal
to block network transmissions for $3$ consecutive time steps. It
is important to note that when the networked control system is periodically
attacked, the specific failure sequence and the period itself are
part of attacker's strategy and in general they are not available
to the system operator. We consider a networked control problem that
covers this case in Section~\ref{sec:Application-to-Networked}.
There, we show that to check networked control system's stability
through Theorem~\ref{Stability-Theorem}, only the knowledge of the
upper-bound on the average attack ratio is needed. For the periodic
attack in Fig.~\ref{Flo:Periodic-Chain}, the upper-bound on this
ratio is given by $\overline{\rho}_{2}$, since the second mode corresponds
to the attacks. Notice that in this case we can select $\underline{\rho}_{1}=1-\overline{\rho}_{2}$,
$\overline{\rho}_{1}=1$, and $\underline{\rho}_{2}=0$.

\begin{figure}
\begin{center}\includegraphics[width=0.85\columnwidth]{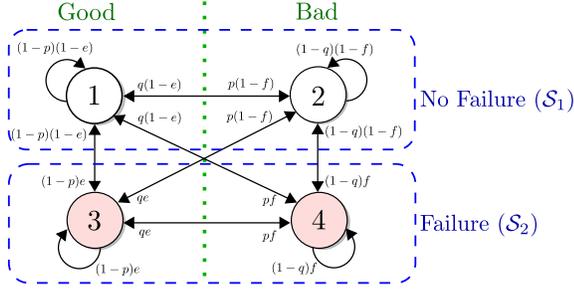}\end{center}
\vskip -5pt

\caption{Transition diagram for Markov chain $\{g(t)\in\mathcal{S}=\{1,2,3,4\}\}_{t\in\mathbb{N}_{0}}$
with partitions $\mathcal{S}_{1}=\{1,2\},\mathcal{S}_{2}=\{3,4\}$,
characterizing packet exchange failures on a Gilbert-Elliott channel.}

\label{Flo:Gilbert-Elliott-Chain} \vskip -12pt
\end{figure}

\emph{Example 2:} The characterization in (\ref{eq:l-g-definition})
can also be used to describe random packet transmission failures.
For example, communication channels following the Markov model can
be described simply by setting $\mathcal{S}_{1}\triangleq\{1\}$,
$\mathcal{S}_{2}\triangleq\{2\}$, and $\{g(t)\in\mathcal{S}=\mathcal{S}_{1}\cup\mathcal{S}_{2}\}_{t\in\mathbb{N}_{0}}$
as a Markov chain with certain transition probabilities. In addition,
the Gilbert-Elliott model and other more advanced models based on
Markov chains (see \cite{sadeghi2008,ellis2014}) can also be described
within the framework. For instance, in Gilbert-Elliott model, the
channel is in the state of either ``Good'' or ``Bad''. In Good
channel state, packet losses occur with a small probability $e$;
moreover, in Bad channel state, failure probability denoted by $f$
may be large. Transitions between Good and Bad states occur with probability
$p$ from Good to Bad and $q$ from Bad to Good. This scenario can
be described by setting $\mathcal{S}_{1}\triangleq\{1,2\}$, $\mathcal{S}_{2}\triangleq\{3,4\}$,
and $\{g(t)\in\mathcal{S}=\mathcal{S}_{1}\cup\mathcal{S}_{2}\}_{t\in\mathbb{N}_{0}}$
as a Markov chain with transition diagram shown in Fig.~\ref{Flo:Gilbert-Elliott-Chain}.
In this setting, $g(t)\in\{1,3\}$ corresponds to Good channel state
and $g(t)\in\{2,4\}$ corresponds to Bad. On the other hand, by (\ref{eq:l-g-definition}),
$g(t)\in\mathcal{S}_{2}$ indicates a packet exchange failure at time
$t$, whereas $g(t)\in\mathcal{S}_{1}$ indicates a successful packet
exchange attempt. Using different settings for $\mathcal{S}_{1}$,
$\mathcal{S}_{2}$, and $\{g(t)\in\mathcal{S}\}_{t\in\mathbb{N}_{0}}$,
we can also model the situation where the network faces both jamming
attacks and random packet transmission failures. 

\begin{inarxiv}

Note that when $\{r(t)\in\mathcal{M}\}_{t\in\mathbb{N}_{0}}$ is characterized
through (\ref{eq:l-g-definition}), the limits $\lim_{k\to\infty}\frac{1}{k}\sum_{i=0}^{k-1}\mathbbm{1}[\bar{r}(i)=q]$,
$q\in\mathcal{M}^{h}$, exist for all $h\in\mathbb{N}$. Hence, in
such cases, the stability analysis in Theorem~\ref{Stability-Theorem}
can be conducted with any $h\in\mathbb{N}$. On the other hand, for
other characterizations of $\{r(t)\in\mathcal{M}\}_{t\in\mathbb{N}_{0}}$,
it may be the case that the limits exist for $h\in\{1,2,\ldots,\hat{h}\}$
but not for $h>\hat{h}$, where $\hat{h}\in\mathbb{N}$. In those
situations, Theorem~\ref{Stability-Theorem} is applicable only for
$h\in\{1,2,\ldots,\hat{h}\}$. 

\end{inarxiv}

\begin{nopf}

In the characterization of Proposition~\ref{Proposition-l-g}, the
initial values $g(0)$ and $r(0)$ are chosen to be deterministic
for simplicity of presentation. Randomness in the initial mode can
be introduced through the following extension. Let $\{r_{j}(t)\in\mathcal{M}\}_{t\in\mathbb{N}_{0}}$,
$j\in\{1,\ldots,N\}$, be finite-state stochastic processes and let
$\{r(t)\in\mathcal{M}\}_{t\in\mathbb{N}_{0}}$ be given by $r(t)=r_{a}(t)$,
$t\in\mathbb{N}_{0}$, where $a\colon\Omega\to\{1,\ldots,N\}$ is
a random variable. Now, for a given $h\in\mathbb{N}$, let $\{\bar{r}_{j}(i)\in\mathcal{M}^{h}\}_{i\in\mathbb{N}_{0}}$,
$j\in\{1,\ldots,N\}$, be sequence-valued processes given by 
\begin{align}
\bar{r}_{j}(i) & \triangleq(r_{j}(ih),r_{j}(ih+1),\ldots,r_{j}((i+1)h-1)),\label{eq:lbar-definition-1}
\end{align}
 for $i\in\mathbb{N}_{0}$. If for each $j\in\{1,\ldots,N\}$, the
limits $\lim_{k\to\infty}\frac{1}{k}\sum_{i=0}^{k-1}\mathbbm{1}[\bar{r}_{j}(i)=q]$,
$q\in\mathcal{M}^{h}$, exist, then the limits $\lim_{k\to\infty}\frac{1}{k}\sum_{i=0}^{k-1}\mathbbm{1}[\bar{r}(i)=q]$,
$q\in\mathcal{M}^{h}$, also exist, since 
\begin{align}
 & \lim_{k\to\infty}\frac{1}{k}\sum_{i=0}^{k-1}\mathbbm{1}[\bar{r}(i)=q]\nonumber \\
 & \quad=\sum_{j=1}^{N}\mathbbm{1}[a=j]\lim_{k\to\infty}\frac{1}{k}\sum_{i=0}^{k-1}\mathbbm{1}[\bar{r}_{j}(i)=q],\quad q\in\mathcal{M}^{h}.\label{eq:a-eq}
\end{align}
 Notice that in this case, the initial mode and the mode signal trajectory
depends on $a$, and thus, $\lim_{k\to\infty}\frac{1}{k}\sum_{i=0}^{k-1}\mathbbm{1}[\bar{r}(i)=q]$
is a random variable that depends on the value of $a$. 

\end{nopf}

\section{Stability Assessment via Linear Programming \label{sec:Linear-Programming-Methods}}

In this section, we investigate two closely-related linear programming
problems and present a method for checking the almost sure asymptotic
stability condition given in Theorem~\ref{Stability-Theorem} through
their optimal solutions. 

\subsection{Linear Programming Problem 1}

Theorem~\ref{Stability-Theorem} states that the switched system
(\ref{eq:switched-system}) is stable if there exists an induced matrix
norm $\|\cdot\|$ and a scalar $\varepsilon\in(0,1)$ such that the
inequality (\ref{eq:maxcondition}) holds for all $\rho_{q}\in[0,1],q\in\mathcal{M}^{h}$,
that satisfy (\ref{eq:rhoqineq1}), (\ref{eq:rhoqineq2}). In what
follows, we provide a linear programming problem to check this condition
for a given induced matrix norm $\|\cdot\|$ and scalar $\varepsilon\in(0,1)$. 

Now define $\gamma_{q}\in\mathbb{R}$, $q\in\mathcal{M}^{h}$, as
in (\ref{eq:small-gamma-definition}) and consider the linear programming
problem 
\begin{align}
 & \begin{array}{cc}
\underset{\rho_{q}\in[0,1],\,q\in\mathcal{M}^{h}}{\text{{maximize}}} & \sum_{q\in\mathcal{M}^{h}}\gamma_{q}\rho_{q}\\
\text{{subject\,to}} & \eqref{eq:rhoqineq1},\eqref{eq:rhoqineq2}.
\end{array}\label{eq:linear-programming-problem-1}
\end{align}

For the stability analysis, different values of $\rho_{q},q\in\mathcal{M}^{h}$,
that satisfy (\ref{eq:rhoqineq1}), (\ref{eq:rhoqineq2}) represent
possible mode activity scenarios such that the long run average conditions
(\ref{eq:infcond}) and (\ref{eq:supcond}) hold. The linear programming
problem (\ref{eq:linear-programming-problem-1}) allows us to identify
scenarios that maximize $\sum_{q\in\mathcal{M}^{h}}\gamma_{q}\rho_{q}$.
We can then check the stability condition (\ref{eq:maxcondition})
with the maximum value of $\sum_{q\in\mathcal{M}^{h}}\gamma_{q}\rho_{q}$
instead of checking it for all possible scenarios.

In the following lemma, we show that the linear programming problem
(\ref{eq:linear-programming-problem-1}) is feasible, that is, there
always exist $\rho_{q}\in[0,1],q\in\mathcal{M}^{h}$, that satisfy
(\ref{eq:rhoqineq1}), (\ref{eq:rhoqineq2}). Furthermore, we show
that the problem is bounded (i.e., the objective function $\sum_{q\in\mathcal{M}^{h}}\gamma_{q}\rho_{q}$
in (\ref{eq:linear-programming-problem-1}) is bounded). 

\begin{lemma}\label{Linear-Programming-Lemma} The linear programming
problem (\ref{eq:linear-programming-problem-1}) is feasible and bounded.\end{lemma}

\begin{seearxiv}

\begin{IEEEproof}First, we show that the feasible region of the linear
programming problem is not empty. To this end, first observe that
$\sum_{s=1}^{M}\underline{\rho}_{s}\leq1\leq\sum_{s=1}^{M}\overline{\rho}_{s}$.
This is because Assumption~\ref{MainAssumption-1} implies 
\begin{align*}
1 & =\limsup_{k\to\infty}\frac{1}{k}\sum_{t=0}^{k-1}\sum_{s=1}^{M}\mathbbm{1}[r(t)=s]\\
 & \leq\sum_{s=1}^{M}\limsup_{k\to\infty}\frac{1}{k}\sum_{t=0}^{k-1}\mathbbm{1}[r(t)=s]\leq\sum_{s=1}^{M}\overline{\rho}_{s},
\end{align*}
 and similarly,
\begin{align*}
1 & =\liminf_{k\to\infty}\frac{1}{k}\sum_{t=0}^{k-1}\sum_{s=1}^{M}\mathbbm{1}[r(t)=s]\\
 & \geq\sum_{s=1}^{M}\limsup_{k\to\infty}\frac{1}{k}\sum_{t=0}^{k-1}\mathbbm{1}[r(t)=s]\geq\sum_{s=1}^{M}\underline{\rho}_{s}.
\end{align*}
Now let $\underline{\varrho}\triangleq\sum_{s=1}^{M}\underline{\rho}_{s}$,
$\overline{\varrho}\triangleq\sum_{s=1}^{M}\overline{\rho}_{s}$ and
$\beta_{s}\triangleq(\overline{\rho}_{s}-\underline{\rho}_{s})\frac{1-\underline{\varrho}}{\overline{\varrho}-\underline{\varrho}}$
and consider $\rho_{q},q\in\mathcal{M}^{h}$, given by 
\begin{align}
\rho_{q} & =\begin{cases}
\underline{\rho}_{s}+\beta_{s},\quad & \mathrm{if}\,\,c_{s}(q)=h,\,\,s\in\mathcal{M},\\
0,\quad & \mathrm{otherwise}.
\end{cases}\label{eq:rhoqinlemmaproof}
\end{align}
To establish that the feasible region contains $\rho_{q},q\in\mathcal{M}^{h}$,
given by (\ref{eq:rhoqinlemmaproof}), we show that (\ref{eq:rhoqineq1})
and (\ref{eq:rhoqineq2}) hold. First, (\ref{eq:rhoqineq1}) holds
because 
\begin{align}
\sum_{q\in\mathcal{M}^{h}}\rho_{q} & =\sum_{s=1}^{M}(\underline{\rho}_{s}+\beta_{s})=\underline{\varrho}+\frac{1-\underline{\varrho}}{\overline{\varrho}-\underline{\varrho}}\sum_{s=1}^{M}(\overline{\rho}_{s}-\underline{\rho}_{s})\nonumber \\
 & =\underline{\varrho}\left(1-\frac{1-\underline{\varrho}}{\overline{\varrho}-\underline{\varrho}}\right)+\frac{1-\underline{\varrho}}{\overline{\varrho}-\underline{\varrho}}\overline{\varrho}=1.\label{eq:prhoqexamplesumisone}
\end{align}
Now we show (\ref{eq:rhoqineq2}). Since $\underline{\varrho}\leq1\leq\overline{\varrho}$,
we have $\beta_{s}\in[0,\overline{\rho}_{s}-\underline{\rho}_{s}]$.
It then follows that 
\begin{align}
\sum_{q\in\mathcal{M}^{h}}c_{s}(q)\rho_{q} & =\underline{\rho}_{s}+\beta_{s}\in[\underline{\rho}_{s},\overline{\rho}_{s}],\quad s\in\mathcal{M}.\label{eq:rhoqexamplesatisfiesbounds}
\end{align}

It now remains to show that the solutions to the linear programming
problem are bounded. Note that since $\rho_{q}\leq1$, $q\in\mathcal{M}^{h}$,
we have $\sum_{q\in\mathcal{M}^{h}}\gamma_{q}\rho_{q}\leq(\max_{q\in\mathcal{M}^{h}}\gamma_{q})\sum_{q\in\mathcal{M}^{h}}\rho_{q}\leq(\max_{q\in\mathcal{M}^{h}}\gamma_{q})M^{h}<\infty$,
which completes the proof. \end{IEEEproof}

\end{seearxiv}

Lemma~\ref{Linear-Programming-Lemma} implies that there exists an
optimal solution to the linear programming problem (\ref{eq:linear-programming-problem-1})
(see Proposition~3.1 of \cite{korte2002combinatorial}). Even though
there may be multiple optimal solutions, we can always compute the
optimal value of the objective function using any one of those solutions.
Let $J_{h}$ denote the optimal value of the objective function when
$h$-length sequences are considered, that is, 
\begin{align*}
J_{h} & \triangleq\max\Big\{\sum_{q\in\mathcal{M}^{h}}\gamma_{q}\rho_{q}\colon\rho_{q}\in[0,1],q\in\mathcal{M}^{h},\,\text{s.t.}\,\eqref{eq:rhoqineq1},\eqref{eq:rhoqineq2}\Big\}.
\end{align*}
The stability of the switched system (\ref{eq:switched-system}) can
be assessed by checking the sign of the optimal value $J_{h}$. Specifically,
the zero solution $x(t)\equiv0$ of the switched system (\ref{eq:switched-system})
is asymptotically stable almost surely  if 
\begin{align}
J_{h} & <0.\label{eq:jhcondition}
\end{align}
This is because (\ref{eq:jhcondition}) implies that (\ref{eq:maxcondition})
in Theorem~\ref{Stability-Theorem} holds for all $\rho_{q}\in[0,1],q\in\mathcal{M}^{h}$,
that satisfy (\ref{eq:rhoqineq1}), (\ref{eq:rhoqineq2}). 

\begin{remark} \label{Remark-Jh-h-issue} The optimal solution $J_{h}$
for the linear programming problem (\ref{eq:linear-programming-problem-1})
may be positive when $h$ is small, but may become negative for sufficiently
large $h$. The reason is that with large $h$, stability/instability
properties of more mode activity patterns are taken into account.
For instance, consider a switched system with two modes. When $h=2$,
the effects of the dynamics associated with packet failure sequences
in $\mathcal{M}^{2}\triangleq\{(1,1),(1,2),(2,1),(2,2)\}$ are represented
by $\gamma_{q},q\in\mathcal{M}^{2}$. However, $\gamma_{q},q\in\mathcal{M}^{2}$,
cannot be used to distinguish the difference between stabilizing (destabilizing)
effects of longer mode activity sequences $(1,2,2,1)$ and $(2,1,1,2)$,
since both of them are composed of the same smaller sequences $(1,2)$,
$(2,1)$. Thus to show stability, we may need to take into account
longer mode sequences and obtain $J_{h}$ for larger values of $h\in\mathbb{N}$.
Note that $\gamma_{q}$ associated with a mode sequence $q=(1,2,2,1)$
may be negative, even though $\gamma_{(1,2)}$ and $\gamma_{(2,1)}$
associated with the smaller sequences $(1,2)$, $(2,1)$ are positive.
This is similar to the observation that a switched system with individually
unstable modes may be stable if the switching is constrained in a
certain way (see Chapter~2 of \cite{ajlp_liberzon2003}). We note
that our approach of choosing a larger $h$ value is related to the
approach utilized in \cite{bolzern2004almost} for showing average
contractivity of Markov jump systems over $h$ steps as well as the
approaches used in \cite{ajlp_huang2011set,lee2006uniform,ajlp_constrained_philippe2016stability_constrained}
for showing switched system stability through the investigation of
the evolution of the mode signal over multiple time steps. In addition
to these works, a related approach was also used in \cite{levy2016contraction,zorzi2015convergence}
to show convergence of risk-sensitive and risk sensitive like filters
by means of establishing strict contractivity of Riccati recursions
over $h$ steps. \end{remark}

Even though there are efficient algorithms for solving linear programming
problems, it is difficult to solve (\ref{eq:linear-programming-problem-1})
and obtain $J_{h}$ when $h\in\mathbb{N}$ is large. This is because
the number of variables $\rho_{q},q\in\mathcal{M}^{h}$, of the problem
(\ref{eq:linear-programming-problem-1}) grows exponentially in $h$.
Specifically, the number of elements of the set $\mathcal{M}^{h}$,
and hence the number of variables of the linear programming problem
(\ref{eq:linear-programming-problem-1}) is given by $f(h,M)\triangleq M^{h}$.
In the following, we show that an alternative linear programming problem
with fewer variables shares the same optimal objective function value
as that of (\ref{eq:linear-programming-problem-1}). In particular,
the number of variables in this alternative problem grows only polynomially
in $h$. 

\subsection{Linear Programming Problem 2}

We observe in the linear programming problem (\ref{eq:linear-programming-problem-1})
that due to the particular structure of our problem setting, some
of the variables have the same coefficients in the constraints. In
particular, if two (or more) mode sequences are reordered versions
of each other, then the variables associated with those mode sequences
have the same coefficients in the constraints. This allows us to obtain
an alternative problem with fewer variables. 

For an intuitive explanation of how this is possible, consider the
case with $M=2$, $h=2$. In this case, the constraints in (\ref{eq:rhoqineq2})
are $\underline{\rho}_{1}\leq\rho_{(1,1)}+0.5\rho_{(1,2)}+0.5\rho_{(2,1)}+0\rho_{(2,2)}\leq\overline{\rho}_{1}$
and $\underline{\rho}_{2}\leq0\rho_{(1,1)}+0.5\rho_{(1,2)}+0.5\rho_{(2,1)}+1\rho_{(2,2)}\leq\overline{\rho}_{2}$,
where the variables $\rho_{(1,2)}$ and $\rho_{(2,1)}$ have the same
coefficients. If, for example, $\gamma_{(2,1)}\geq\gamma_{(1,2)}$,
then it means that we can maximize the objective $\gamma_{(1,1)}\rho_{(1,1)}+\gamma_{(1,2)}\rho_{(1,2)}+\gamma_{(2,1)}\rho_{(2,1)}+\gamma_{(2,2)}\rho_{(2,2)}$
by setting $\rho_{(1,2)}$ to $0$ and considering only $\rho_{(1,1)}$,
$\rho_{(2,1)}$, $\rho_{(2,2)}$ as variables of the optimization
problem. This is because if $\rho_{q}=\tilde{\rho}_{q}$, $q\in\mathcal{M}^{h}$,
is an optimal solution of the problem, then $\rho_{q}=\hat{\rho}_{q},q\in\mathcal{M}^{h}$,
with $\hat{\rho}_{(1,1)}=\tilde{\rho}_{(1,1)}$, $\hat{\rho}_{(1,2)}=0$,
$\hat{\rho}_{(2,1)}=\tilde{\rho}_{(1,2)}+\tilde{\rho}_{(2,1)}$, and
$\hat{\rho}_{(2,2)}=\tilde{\rho}_{(2,2)}$ is also an optimal solution
since $\rho_{q}=\hat{\rho}_{q},q\in\mathcal{M}^{h},$ satisfy the
constraints and $\gamma_{(2,1)}\geq\gamma_{(1,2)}$ implies $\sum_{q\in\mathcal{M}^{h}}\gamma_{q}\tilde{\rho}_{q}\leq\sum_{q\in\mathcal{M}^{h}}\gamma_{q}\hat{\rho}_{q}$.
(Furthermore, if $\gamma_{(2,1)}$ is strictly larger than $\gamma_{(1,2)}$,
then for all optimal solutions $\rho_{q}$, $q\in\mathcal{M}^{h}$,
the value of $\rho_{(1,2)}$ is necessarily $0$.) As this discussion
indicates, the variable $\rho_{(1,2)}$ can be removed from the optimization
problem for this example, since the terms that involve $\rho_{(1,2)}$
are all $0$. Similarly, if $\gamma_{(2,1)}\leq\gamma_{(1,2)}$, then
the variable $\rho_{(2,1)}$ can be removed. Notice that similar techniques
can be used for all $M\in\mathbb{N}$ and $h\in\mathbb{N}$.  Furthermore,
more variable reductions are achieved as $M$ and $h$ increase. In
particular, we obtain the following linear programming problem:

Let $\mathcal{Z}_{h}\triangleq\{(z_{1},z_{2},\ldots,z_{M})\colon z_{s}\in\{0,1,\ldots,h\},s\in\mathcal{M},{\textstyle \sum}_{s=1}^{M}z_{s}=h\}$,
and consider the linear programming problem 
\begin{align}
 & \begin{array}{cc}
\underset{\rho_{z}'\in[0,1],\,z\in\mathcal{Z}_{h}}{\text{{maximize}}} & \sum_{z\in\mathcal{Z}_{h}}\gamma_{z}'\rho_{z}'\qquad\qquad\\
\text{{subject\,to}} & \sum_{z\in\mathcal{Z}_{h}}\rho_{z}'=1,\qquad\,\qquad\qquad\,\,\,\,\,\,\,(a)\\
 & \underline{\rho}_{s}\leq\sum_{z\in\mathcal{Z}_{h}}\frac{z_{s}}{h}\rho'_{z}\leq\overline{\rho}_{s},\,s\in\mathcal{M},\,(b)
\end{array}\label{eq:linear-programming-problem-2}
\end{align}
 where 
\begin{align}
\gamma{}_{z}' & \triangleq\max_{q\in\mathcal{M}^{h,z}}\gamma_{q},\quad z\in\mathcal{Z}_{h},\label{eq:lpp2-small-gamma-i-definition}
\end{align}
with $\mathcal{M}^{h,z}\triangleq\{q\in\mathcal{M}^{h}\,\colon\,c_{1}(q)=z_{1},\,c_{2}(q)=z_{2},c_{3}(q)=z_{3}\,\ldots,c_{M}(q)=z_{M}\}$,
$z\in\mathcal{Z}_{h}$. 

In what follows, we first show that the objective functions of the
linear programming problems (\ref{eq:linear-programming-problem-1})
and (\ref{eq:linear-programming-problem-2}) have the same optimal
values. After that we discuss the advantage of the linear programming
problem (\ref{eq:linear-programming-problem-2}) over the problem
(\ref{eq:linear-programming-problem-1}). \begin{inarxiv}Specifically,
we show that it is easier to solve the linear programming problem
(\ref{eq:linear-programming-problem-2}) because it involves fewer
variables. \end{inarxiv}

\begin{lemma}\label{Linear-Programming-Lemma-2} The linear programming
problem (\ref{eq:linear-programming-problem-2}) is feasible and bounded.\end{lemma}

\begin{seearxiv}

\begin{IEEEproof} The proof is similar to that of Lemma~\ref{Linear-Programming-Lemma}.
First, we show that the feasible region of the linear programming
problem (\ref{eq:linear-programming-problem-2}) is not empty. To
this end, consider $\rho_{z}',\,z\in\mathcal{Z}_{h}$, given by 
\begin{align}
\rho_{z}' & =\sum_{q\in\mathcal{M}^{h,z}}\rho_{q},\label{eq:rhozinexample}
\end{align}
 with $\rho_{q}$ given in (\ref{eq:rhoqinlemmaproof}). Notice that
$\rho_{z}'\in[0,1],\,z\in\mathcal{Z}_{h}$. This is because $\rho_{q}\geq0$,
$q\in\mathcal{M}^{h}$, implies $\rho_{z}'\geq0$, and moreover, $\rho_{z}'=\sum_{q\in\mathcal{M}^{h,z}}\rho_{q}\leq\sum_{q\in\mathcal{M}^{h}}\rho_{q}=1$
due to (\ref{eq:prhoqexamplesumisone}). To establish that the feasible
region contains $\rho_{z}',z\in\mathcal{Z}_{h}$, given by (\ref{eq:rhozinexample}),
we show that (\ref{eq:linear-programming-problem-2}a) and (\ref{eq:linear-programming-problem-2}b)
hold. Now, (\ref{eq:linear-programming-problem-2}a) holds, because
it follows from (\ref{eq:prhoqexamplesumisone}) that 
\begin{align*}
\sum_{z\in\mathcal{Z}_{h}}\rho_{z}' & =\sum_{z\in\mathcal{Z}_{h}}\sum_{q\in\mathcal{M}^{h,z}}\rho_{q}=\sum_{q\in\mathcal{M}^{h}}\rho_{q}=1,
\end{align*}
 where we also used the fact that $\mathcal{M}^{h,z}\cap\mathcal{M}^{h,\hat{z}}=\emptyset$
for $z\neq\hat{z}$, $z,\hat{z}\in\mathcal{Z}_{h}$, and $\cup_{z\in\mathcal{Z}_{h}}\mathcal{M}^{h,z}=\mathcal{M}^{h}$. 

Next, to show (\ref{eq:linear-programming-problem-2}b), note that
$\sum_{z\in\mathcal{Z}_{h}}\frac{z_{s}}{h}\rho'_{z}=\sum_{z\in\mathcal{Z}_{h}}\sum_{q\in\mathcal{M}^{h,z}}\frac{z_{s}}{h}\rho_{q}=\sum_{q\in\mathcal{M}^{h}}\frac{c_{s}(q)}{h}\rho_{q}$.
Hence, (\ref{eq:rhoqexamplesatisfiesbounds}) implies (\ref{eq:linear-programming-problem-2}b). 

It now remains to show that the solutions to the linear programming
problem are bounded. Note that since $\rho_{z}'\leq1$, $z\in\mathcal{Z}_{h}$,
it follows from (\ref{eq:lpp2-small-gamma-i-definition}) that $\sum_{z\in\mathcal{Z}_{h}}\gamma_{z}'\rho_{z}'\leq(\max_{z\in\mathcal{Z}_{h}}\gamma_{z}')\sum_{z\in\mathcal{Z}_{h}}\rho_{z}'\leq(\max_{q\in\mathcal{M}^{h}}\gamma_{q})\binom{h+M-1}{M-1}<\infty$,
which completes the proof. \end{IEEEproof}

\end{seearxiv}

By Lemma~\ref{Linear-Programming-Lemma-2}, there exists an optimal
solution to the linear programming problem (\ref{eq:linear-programming-problem-2}).
Let $J'_{h}$ denote the optimal value of the objective function for
a given $h\in\mathbb{N}$, that is, 
\begin{align*}
J'_{h} & \triangleq\max\Big\{\sum_{z\in\mathcal{Z}_{h}}\gamma_{z}'\rho_{z}'\colon\rho_{z}',z\in\mathcal{Z}_{h},\,\text{s.t.}\,(\ref{eq:linear-programming-problem-2}a),(\ref{eq:linear-programming-problem-2}b)\Big\}.
\end{align*}
 \begin{lemma} \label{Lemma-equivalence} The objective functions
of the linear programming problems (\ref{eq:linear-programming-problem-1})
and (\ref{eq:linear-programming-problem-2}) have the same optimal
values, that is, $J_{h}=J'_{h}$. \end{lemma} 

\begin{IEEEproof} We prove this result by showing $J_{h}\leq J'_{h}$
and $J_{h}\geq J'_{h}$ separately. 

To establish $J_{h}\leq J'_{h}$, we show that for all $\rho_{q}\in[0,1],q\in\mathcal{M}^{h},$
such that (\ref{eq:rhoqineq1}), (\ref{eq:rhoqineq2}) hold, we have
$\sum_{q\in\mathcal{M}^{h}}\gamma_{q}\rho_{q}\leq J'_{h}$. Now, notice
that for all $\rho_{q}\in[0,1],q\in\mathcal{M}^{h}$, such that (\ref{eq:rhoqineq1}),
(\ref{eq:rhoqineq2}) hold, the constraints in (\ref{eq:linear-programming-problem-2}a)
and (\ref{eq:linear-programming-problem-2}b) hold with $\rho_{z}'\triangleq\sum_{q\in\mathcal{M}^{h,z}}\rho_{q}$,
$z\in\mathcal{Z}_{h}$, and as a result, $\sum_{z\in\mathcal{Z}_{h}}\gamma_{z}'\rho_{z}'\leq J'_{h}$.
Hence, for all $\rho_{q},q\in\mathcal{M}^{h}$, such that (\ref{eq:rhoqineq1}),
(\ref{eq:rhoqineq2}) hold, we have 
\begin{align*}
\sum_{q\in\mathcal{M}^{h}}\gamma_{q}\rho_{q} & =\sum_{z\in\mathcal{Z}_{h}}\sum_{q\in\mathcal{M}^{h,z}}\gamma_{q}\rho_{q}\leq\sum_{z\in\mathcal{Z}_{h}}\sum_{q\in\mathcal{M}^{h,z}}\gamma_{z}'\rho_{q}\\
 & =\sum_{z\in\mathcal{Z}_{h}}\gamma_{z}'\sum_{q\in\mathcal{M}^{h,z}}\rho_{q}=\sum_{z\in\mathcal{Z}_{h}}\gamma_{z}'\rho_{z}'\leq J'_{h},
\end{align*}
which implies $J_{h}\leq J'_{h}$. 

To prove $J_{h}\geq J'_{h}$, we now show that there exists $\rho_{q}\in[0,1],q\in\mathcal{M}^{h}$,
such that $\sum_{q\in\mathcal{M}^{h}}\gamma_{q}\rho_{q}=J'_{h}$ and
(\ref{eq:rhoqineq1}), (\ref{eq:rhoqineq2}) hold. Here, let $\hat{\rho}'_{z},z\in\mathcal{Z}_{h}$,
denote an optimal solution to the linear programming problem (\ref{eq:linear-programming-problem-2}),
that is, 
\begin{align*}
\sum_{i\in\mathcal{Z}_{h}}\gamma_{z}'\hat{\rho}'_{z} & =J'_{h}.
\end{align*}
Now, for each $z\in\mathcal{Z}_{h}$, let $q^{(z)}\in\mathrm{argmax}_{q\in\mathcal{M}^{h,z}}\gamma_{q}$
and set $\rho_{q^{(z)}}=\hat{\rho}'_{z}$, $\rho_{q}=0$, $q\in\mathcal{M}^{h,z}\setminus\{q^{(z)}\}$.
It follows that $\rho_{q},q\in\mathcal{M}^{h},$ satisfy (\ref{eq:rhoqineq1}),
(\ref{eq:rhoqineq2}); furthermore, 
\begin{align*}
\sum_{q\in\mathcal{M}^{h}}\gamma_{q}\rho_{q} & =\sum_{z\in\mathcal{Z}_{h}}\gamma_{z}'\hat{\rho}_{z}'=J'_{h}.
\end{align*}
This establishes that there exist $\rho_{q}\in[0,1],q\in\mathcal{M}^{h}$,
such that $\sum_{q\in\mathcal{M}^{h}}\gamma_{q}\rho_{q}=J'_{h}$ and
(\ref{eq:rhoqineq1}), (\ref{eq:rhoqineq2}) hold, which implies that
$J_{h}\geq J'_{h}$. \end{IEEEproof} 

A direct consequence of Lemma~\ref{Lemma-equivalence} is that if
$J'_{h}<0$, then the zero solution $x(t)\equiv0$ of the switched
system (\ref{eq:switched-system}) is asymptotically stable almost
surely. 

\begin{inarxiv}So far we established that almost sure asymptotic
stability of the switched system (\ref{eq:switched-system}) can be
assessed by checking signs of the optimal objective function values
($J_{h}$ and $J'_{h}$) of linear programming problems (\ref{eq:linear-programming-problem-1})
and (\ref{eq:linear-programming-problem-2}). The switched system
(\ref{eq:switched-system}) is stable if the value of $J_{h}=J'_{h}$
is negative. \end{inarxiv} 

\begin{figure}
\vskip 6pt \centering  \includegraphics[width=0.85\columnwidth]{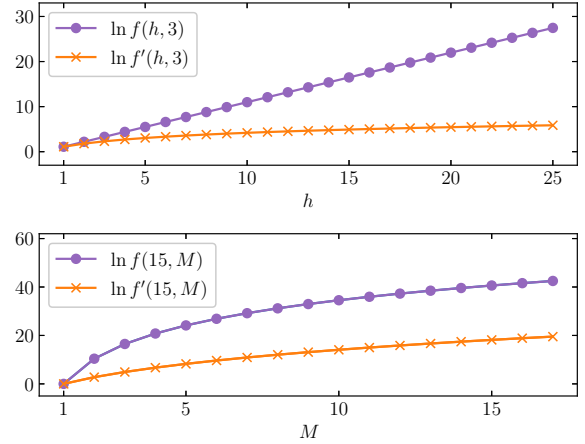}
\vskip -5pt

\caption{Comparison of $\ln f(h,M)$ and $\ln f'(h,M)$ (Top: varying $h$,
fixed $M=3$; Bottom: fixed $h=15$, varying $M$). }
 \label{Flo:lp-comparison} \vskip -12pt
\end{figure}

We observe that solving the linear programming problem (\ref{eq:linear-programming-problem-2})
can be computationally more advantageous in comparison to the problem
(\ref{eq:linear-programming-problem-1}). This is because (\ref{eq:linear-programming-problem-2})
involves fewer variables. Specifically, the number of elements of
the set $\mathcal{Z}_{h}$, and hence the number of variables of the
linear programming problem (\ref{eq:linear-programming-problem-2}),
is $f'(h,M)\triangleq\binom{h+M-1}{M-1}=\frac{(h+M-1)!}{(M-1)!h!}$.
For $h$ and $M$ larger than $1$, the number of variables in the
problem (\ref{eq:linear-programming-problem-2}) is strictly smaller
than that of the problem (\ref{eq:linear-programming-problem-1}),
that is, 
\begin{align*}
f'(h,M) & <f(h,M),\quad h>1,\quad M>1.
\end{align*}
We note again that $f$ grows \emph{exponentially} in $h$. On the
other hand $f'$, the number of variables in the problem (\ref{eq:linear-programming-problem-2}),
grows only \emph{polynomially }in $h$. Specifically, we have 
\begin{align*}
f'(\alpha h,M) & \leq\alpha^{M-1}f'(h,M),\quad\alpha,h,M\in\mathbb{N}.
\end{align*}
 As a result, when $h$ is large, obtaining $J'_{h}$ is much faster
than obtaining $J_{h}$. We also remark that the number of variables
grows polynomially in the number of modes $M$ as well. In particular,
we have
\begin{align*}
f'(h,\alpha M) & \leq\alpha^{h}f'(h,M),\quad\alpha,h,M\in\mathbb{N}.
\end{align*}
 Fig.~\ref{Flo:lp-comparison} shows graphs of $\ln f(h,M)$ and
$\ln f'(h,M)$ indicating how the numbers of variables in problems
(\ref{eq:linear-programming-problem-1}) and (\ref{eq:linear-programming-problem-2})
grow with respect to $h$ and $M$. The difference becomes larger
as $M$ and $h$ get larger. For instance, in the case where $h=15$
and $M=3$, there are $f(15,3)=14{,}348{,}907$ variables in the problem
(\ref{eq:linear-programming-problem-1}), whereas (\ref{eq:linear-programming-problem-2})
involves only $f'(15,3)=136$ variables. 

Notice that an optimal solution $\hat{\rho}'_{z},z\in\mathcal{Z}_{h}$,
to the linear programming problem (\ref{eq:linear-programming-problem-2})
can be used to obtain an optimal solution  $\hat{\rho}{}_{q},q\in\mathcal{M}^{h}$,
to the problem (\ref{eq:linear-programming-problem-1}) by selecting
$q^{(z)}\in\mathrm{argmax}_{q\in\mathcal{M}^{h,z}}\gamma_{q}$ for
each $z\in\mathcal{Z}_{h}$ and then setting $\hat{\rho}{}_{q},q\in\mathcal{M}^{h}$,
as $\hat{\rho}_{q^{(z)}}=\hat{\rho}'_{z}$, $\hat{\rho}_{q}=0$, $q\in\mathcal{M}^{h,z}\setminus\{q^{(z)}\}$,
$z\in\mathcal{Z}_{h}$. As established in the second part of the proof
of Lemma~\ref{Lemma-equivalence}, $\hat{\rho}{}_{q},q\in\mathcal{M}^{h}$,
satisfy $\sum_{q\in\mathcal{M}^{h}}\gamma_{q}\hat{\rho}_{q}=J'_{h}$,
and hence provide an optimal solution to the problem (\ref{eq:linear-programming-problem-1}),
since $J_{h}=J'_{h}$. Clearly, if for any $z\in\mathcal{Z}_{h}$
the set $\mathrm{argmax}_{q\in\mathcal{M}^{h,z}}\gamma_{q}$ has more
than one element, then there are multiple optimal solutions to the
problem (\ref{eq:linear-programming-problem-1}). 

It is important to note that although solving the linear programming
problem (\ref{eq:linear-programming-problem-2}) is easier due to
fewer variables, it requires precomputation of coefficients $\gamma'_{z}$
by (\ref{eq:lpp2-small-gamma-i-definition}). Notice that, for each
$z$, the complexity of computing $\gamma'_{z}$ is linear in the
number of variables of the set $\mathcal{M}^{h,z}$, which is given
by $\frac{h!}{z_{1}!z_{2}!\cdots z_{M}!}$. It turns out that this
computation can be carried out for all coefficients $\gamma_{z}'$
at the same time using parallel computing techniques. For instance,
in the case of $h=15$ and $M=3$, the set $\mathcal{Z}_{h}$ of $f'(15,3)=136$
variables can be partitioned into $8$ subsets with size $17$. We
can then utilize a computer with $8$ central processing units to
carry out the computation for each of these subsets. Both linear programming
problems (\ref{eq:linear-programming-problem-1}) and (\ref{eq:linear-programming-problem-2})
require calculation of norms of matrix products in the computation
of $\gamma_{q}$, $q\in\mathcal{M}^{h}$, given in (\ref{eq:small-gamma-definition}).
For large values of $h$, this is a computationally intensive calculation,
but it can also be conducted in parallel for different $q$ values. 

We also note that using different matrix norms in the definition of
$\gamma_{q},q\in\mathcal{M}^{h}$, can be useful to check stability
in the case of limited computational resources. This is because $J_{h}$
and $J_{h}'$ may be positive for a particular matrix norm and negative
for another. Note also that $\gamma_{q},q\in\mathcal{M}^{h}$, depend
on the value of $\varepsilon$ in (\ref{eq:small-gamma-definition})
if $\Gamma_{q}=0$ for some $q\in\mathcal{M}^{h}$. If $\Gamma_{q}=0$
for some $q\in Q_{h}$ and the optimal solution value $J_{h}=J_{h}'$
is positive for a particular value of $\varepsilon$, then we can
try solving the linear programming problems with a smaller $\varepsilon$
value. If the optimal solution value $J_{h}=J_{h}'$ is negative for
a value $\varepsilon=\hat{\varepsilon}$, then the stabilization is
guaranteed. In that case, we do not need to consider smaller $\varepsilon$
values, since for every $\varepsilon\in(0,\hat{\varepsilon})$, the
optimal solution value $J_{h}=J_{h}'$ is also guaranteed to be negative.
This is because the constraints in the linear programming problems
do not change with $\varepsilon$ and the coefficients of the objective
functions for the case with $\varepsilon\in(0,\hat{\varepsilon})$
are smaller than the case with $\varepsilon=\hat{\varepsilon}$. 

\begin{inarxiv} We remark that for stability analysis, the sign of
$J_{h}$ and $J'_{h}$ can also be assessed by solving linear feasibility
problems without computing the actual values of those scalars. In
particular, we have $J_{h}\geq0$ if and only if there exist $\rho_{q}\in[0,1]$,
$q\in\mathcal{M}^{h}$, such that (\ref{eq:rhoqineq1}), (\ref{eq:rhoqineq2}),
and $\sum_{q\in\mathcal{M}^{h}}\gamma_{q}\rho_{q}\geq0$ hold. Similarly,
we have $J'_{h}\geq0$ if and only if there exist $\rho_{z}'\in[0,1]$,
$z\in\mathcal{Z}_{h}$, such that (\ref{eq:linear-programming-problem-2}a),
(\ref{eq:linear-programming-problem-2}b), and $\sum_{z\in\mathcal{Z}_{h}}\gamma_{z}'\rho_{z}'\geq0$
hold. Note that solving these feasibility problems is not necessarily
faster in comparison to solving the associated linear programming
problems, since the numbers of variables in these two feasibility
problems are equal to those in the associated linear programming problems.
\end{inarxiv} 

\section{Application to Networked Control Under Jamming Attacks \label{sec:Application-to-Networked}}

In this section we consider two problem settings where we model the
networked control system as a switched system. 

\subsection{Control over Delay-Free Communication Links \label{subsec:Control-over-Delay-Free}}

\begin{figure}
\vskip 6pt \centering  \includegraphics[width=0.9\columnwidth]{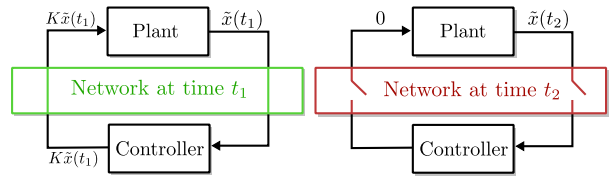}
\vskip -5pt

\caption{Packet exchange success mode (left) and failure mode (right) of the
networked control system.}
 \label{operation} \vskip -12pt
\end{figure}

First, we explore the networked control problem where at each time
instant, the plant and the controller attempt to exchange state and
control input packets over a communication channel. In this problem
setting, network transmissions do not face delay, but packet exchange
attempts between the plant and the controller may be subject to packet
losses due to malicious jamming attacks or nonmalicious communication
errors. In a successful packet exchange attempt, the plant transmits
the state information to the controller; the controller uses the received
state information to compute the control input through a linear control
law and sends back the control input to the plant. The transmitted
control input is then applied at the plant side. Packet exchange attempt
failures happen when either the measured state packets or the control
input packets are lost. In that case, the control input at the plant
side is set to $0$. Fig.~\ref{operation} illustrates the operation
of the networked control system during a successful packet exchange
attempt at time $t_{1}$ and a failed exchange attempt at time $t_{2}$. 

The dynamics of the linear plant is given by 
\begin{align}
\tilde{x}(t+1) & =A\tilde{x}(t)+Bu(t),\quad\tilde{x}(0)=\tilde{x}_{0},\quad t\in\mathbb{N}_{0},\label{eq:system}
\end{align}
 where $\tilde{x}(t)\in\mathbb{R}^{\tilde{n}}$ and $u(t)\in\mathbb{R}^{m}$
denote the state and the control input, respectively; furthermore,
$A\in\mathbb{R}^{\tilde{n}\times\tilde{n}}$ and $B^{\tilde{n}\times m}$
are the state and input matrices, respectively. 

We use the binary-valued process $\{l(t)\in\{0,1\}\}_{t\in\mathbb{N}_{0}}$
to describe success or failure states of packet exchange attempts.
Specifically, the state $l(t)=0$ indicates that the packet exchange
attempt at time $t$ is successful, whereas $l(t)=1$ indicates failure.
In this case, the control input $u(t)$ applied at the plant side
is given by 
\begin{align}
u(t) & \triangleq\left(1-l(t)\right)K\tilde{x}(t),\quad t\in\mathbb{N}_{0},\label{eq:control-input}
\end{align}
where $K\in\mathbb{R}^{m\times\tilde{n}}$ denotes the feedback gain. 

In \cite{cetinkaya2016tac}, we proposed a characterization for $\{l(t)\in\{0,1\}\}_{t\in\mathbb{N}_{0}}$
that allows us to model the effects of random packet losses and jamming
attacks in a unified manner. This characterization relies on the following
assumption. 

\begin{assumption} \label{MainAssumption} There exists a scalar
$\rho\in[0,1]$ such that 
\begin{align}
\sum_{k=1}^{\infty}\mathbb{P}[\sum_{i=0}^{k-1}l(i)>\rho k] & <\infty.\label{eq:lcond1}
\end{align}

\end{assumption} \vskip 10pt

Here, the inequality (\ref{eq:lcond1}) can be considered as a condition
on the evolution of tail probability $\mathbb{P}\big[\sum_{i=0}^{k-1}l(i)/k>\rho\big]$.
The scalar $\rho\in[0,1]$ in (\ref{eq:lcond1}) plays a key role
in characterizing a probabilistic bound on the average ratio of packet
exchange failures. Observe that the case where all packet transmission
attempts result in failure can be described by setting $\rho=1$.
It is shown in \cite{cetinkaya2016tac} that $\rho$ in (\ref{eq:lcond1})
can be obtained to be strictly smaller than $1$ for certain random
and malicious packet loss models. These models include time-inhomogeneous
Markov chains for describing random packet losses, as well as a discrete-time
version of the malicious attack model in \cite{de2015inputtran},
where the number of packet exchange attempts that face attacks is
upper bounded by a certain ratio of the total number of packet exchange
attempts. 

We showed in \cite{cetinkaya2016tac} that the closed-loop networked
control system is stable, when $\rho$ in (\ref{eq:lcond1}) takes
a sufficiently small value. In what follows we provide an alternative
stability analysis method for the networked control system by utilizing
Theorem~\ref{Stability-Theorem}. This new method turns out to be
less conservative than the results in \cite{cetinkaya2016tac} in
certain scenarios. To utilize Theorem~\ref{Stability-Theorem} we
first describe the closed-loop system as a discrete-time switched
system. 

The networked control system (\ref{eq:system}), (\ref{eq:control-input})
is equivalently described as a discrete-time switched system (\ref{eq:switched-system})
with the state $x(t)=\tilde{x}(t)$ and the mode signal given by $r(t)=l(t)+1$.
Furthermore, the subsystem matrices are given by $A_{1}=A+BK$, $A_{2}=A$.
Now, under Assumption~\ref{MainAssumption}, the inequalities (\ref{eq:infcond})
and (\ref{eq:supcond}) in Assumption~\ref{MainAssumption-1} hold
with $\underline{\rho}_{1}=1-\rho,\overline{\rho}_{1}=1,\underline{\rho}_{2}=0$,
and $\overline{\rho}_{2}=\rho$. This is because (\ref{eq:lcond1})
implies 
\begin{align}
\limsup_{k\to\infty}\frac{1}{k}\sum_{t=0}^{k-1}l(t) & \leq\rho,\label{eq:long-run-average-bound}
\end{align}
 and hence $\limsup_{k\to\infty}\frac{1}{k}\sum_{t=0}^{k-1}\mathbbm{1}[r(t)=2]\leq\rho.$
Through this switched system characterization, stability of the networked
control system (\ref{eq:system}), (\ref{eq:control-input}) can be
analyzed by using Theorem~\ref{Stability-Theorem}. Moreover, the
linear programming problems developed in Section~\ref{sec:Linear-Programming-Methods}
can also be employed. 

In \cite{cetinkaya2016tac} (see also \cite{ahmet-cdc}), an event-triggering
controller is used for stabilization, and the packet exchange attempt
times are decided by utilizing a set of triggering conditions. These
conditions can be adjusted to consider the problem setting where the
plant and the controller attempt packet exchanges at each time instant.
For this problem setting, Theorem~\ref{Stability-Theorem} is less
conservative and can be considered as an enhancement of the stability
result presented in \cite{cetinkaya2016tac}. In fact, the stability
result in \cite{cetinkaya2016tac} is obtained by analyzing the evolution
of a Lyapunov-like function $V(x)=x^{\mathrm{T}}Px$ at each time
step. This analysis idea can be recovered by our approach presented
in this paper through setting $h=1$, and defining the norm in (\ref{eq:small-gamma-definition})
as the matrix norm induced by the vector norm $\|x\|_{P}\triangleq\sqrt{x^{\mathrm{T}}Px}$.
In particular, for the case where $\lim_{k\to\infty}\frac{1}{k}\sum_{i=0}^{k-1}l(i)$
exists, for all scenarios in which the stability condition in Theorem~3.5
in \cite{cetinkaya2016tac} holds, the stability condition in Theorem~\ref{Stability-Theorem}
also holds. Furthermore, as we illustrate in Section~\ref{sec:Numerical-Example},
there are cases where the condition in Theorem~3.5 in \cite{cetinkaya2016tac}
does not hold but the condition in Theorem~\ref{Stability-Theorem}
is satisfied. 

\begin{inarxiv} The following result shows that when $\lim_{k\to\infty}\frac{1}{k}\sum_{i=0}^{k-1}l(i)$
exists, for all scenarios in which the stability condition in Theorem~3.5
in \cite{cetinkaya2016tac} holds, the stability condition in Theorem~\ref{Stability-Theorem}
also holds. 

\begin{proposition}Assume $\lim_{k\to\infty}\frac{1}{k}\sum_{i=0}^{k-1}l(i)$
exists. Suppose the stability condition in Theorem~3.5 in \cite{cetinkaya2016tac}
holds, that is, there exist a positive-definite matrix $P\in\mathbb{R}^{n\times n}$
and scalars $\beta\in(0,1)$, $\varphi\in[1,\infty)$ such that 
\begin{align}
 & \beta P-\left(A+BK\right)^{\mathrm{T}}P\left(A+BK\right)\geq0,\label{eq:betacond}\\
 & \varphi P-A^{\mathrm{T}}PA\geq0,\label{eq:varphicond}\\
 & (1-\rho)\ln\beta+\rho\ln\varphi<0,\label{eq:betaandvarphicond}
\end{align}
hold. Then with $h\triangleq1$ the stability condition in Theorem~\ref{Stability-Theorem}
also holds, i.e., $\lim_{k\to\infty}\frac{1}{k}\sum_{i=0}^{k-1}\mathbbm{1}[\bar{r}(i)=q]$,
$q\in\mathcal{M}^{h}$, exist; moreover, there is a matrix norm $\|\cdot\|$
and a scalar $\varepsilon\in(0,1)$ such that (\ref{eq:maxcondition})
holds for all $\rho_{q}\in[0,1]$, $q\in\mathcal{M}^{h}$, that satisfy
(\ref{eq:rhoqineq1}) and (\ref{eq:rhoqineq2}). \end{proposition} 

\begin{IEEEproof} First, note that when $h=1$, we have $\mathcal{M}^{h}=\{(1),(2)\}$.
As a result, existence of $\lim_{k\to\infty}\frac{1}{k}\sum_{i=0}^{k-1}l(i)$
implies that $\lim_{k\to\infty}\frac{1}{k}\sum_{i=0}^{k-1}\mathbbm{1}[\bar{r}(i)=q]$,
$q\in\mathcal{M}^{h}$, also exist.

Next, let the norm in (\ref{eq:small-gamma-definition}) be the matrix
norm induced by the vector norm $\|x\|_{P}\triangleq\sqrt{x^{\mathrm{T}}Px}$,
that is, $\|M\|\triangleq\sup_{x\in\mathbb{R}^{n}\setminus\{0\}}\frac{\|Mx\|_{P}}{\|x\|_{P}},M\in\mathbb{R}^{n\times n}$.
Under this norm, the inequalities (\ref{eq:betacond}) and (\ref{eq:varphicond})
imply $\|A_{1}\|=\|A+BK\|\leq\sqrt{\beta}$ and $\|A_{2}\|=\|A\|\leq\sqrt{\varphi}$.
Now let $\varepsilon\triangleq\sqrt{\beta}$. Since $\varepsilon=\sqrt{\beta}<\sqrt{\varphi}$,
by using (\ref{eq:betaandvarphicond}) and $\rho_{(2)}\leq\overline{\rho}_{2}=\rho$,
we get 
\begin{align*}
 & \sum_{q\in\mathcal{M}^{h}}\gamma_{q}\rho_{q}=\gamma_{(1)}\rho_{(1)}+\gamma_{(2)}\rho_{(2)}=\gamma_{(1)}(1-\rho_{(2)})+\gamma_{(2)}\rho_{(2)}\\
 & \,\,=(1-\rho_{(2)})\ln\max\{\|A+BK\|,\varepsilon\}+\rho_{(2)}\ln\max\{\|A\|,\varepsilon\}\\
 & \,\,\leq(1-\rho_{(2)})\ln\sqrt{\beta}+\rho_{(2)}\ln\sqrt{\varphi}\\
 & \,\,\leq(1-\rho)\ln\sqrt{\beta}+\rho\ln\sqrt{\varphi}=\frac{1}{2}((1-\rho)\ln\beta+\rho\ln\varphi)<0,
\end{align*}
for all $\rho_{q}\in[0,1]$, $q\in\mathcal{M}^{h}=\{(1),(2)\}$, such
that (\ref{eq:rhoqineq1}) and (\ref{eq:rhoqineq2}) hold, which completes
the proof. \end{IEEEproof}

\end{inarxiv}

\subsection{Control over Delay-Free and One-Step Delayed Communication Links
\label{subsec:Control-over-Delayed}}

\begin{figure}
\vskip 6pt \centering  \includegraphics[width=0.6\columnwidth]{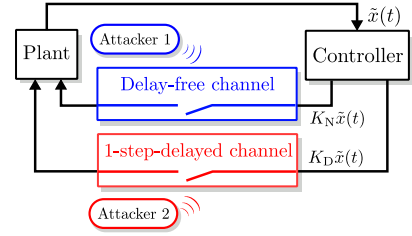}
\vskip -5pt

\caption{Networked control over delay-free and 1-step-delayed channels.}
 \label{operation-2} \vskip -12pt
\end{figure}

The switched system framework generalizes \cite{cetinkaya2016tac}
to the multiple mode case with $M>2$. This aspect is now illustrated
through the networked control system depicted in Fig.~\ref{operation-2},
where the control actions are transmitted to the plant over two separate
communication channels. We assume that one of these channels faces
no delay and the other one faces $1$-step delay in transmissions.
Investigation of a networked control setup involving multiple channels
with different delays is useful for analyzing systems that incorporate
multiple actuators placed at different locations. The nodes that relay
the information coming from the controller to certain actuators may
induce delays due to different security measures in transmission powers,
encryptions, and so on. 

In our problem setting, the plant is as given in (\ref{eq:system}).
The controller receives the system state $\tilde{x}(t)$ at each time
$t$, and computes two control inputs $K_{\mathrm{N}}\tilde{x}(t)$
and $K_{\mathrm{D}}\tilde{x}(t)$ that are attempted to be transmitted
on the delay-free and $1$-step-delayed channels, respectively. We
respectively use $\{l_{\mathrm{N}}(t)\in\{0,1\}\}_{t\in\mathbb{N}_{0}}$
and $\{l_{\mathrm{D}}(t)\in\{0,1\}\}_{t\in\mathbb{N}_{0}}$ to indicate
failures on the delay-free and the one-step-delayed channels. If both
channels fail ($l_{\mathrm{N}}(t)=1$, $l_{\mathrm{D}}(t)=1$), the
control input at the plant side is set to $0$. Furthermore, the control
data $K_{\mathrm{D}}\tilde{x}(t-1)$ received from the delayed channel
is used only if the transmission on the delay-free channel fails ($l_{\mathrm{N}}(t)=1$,
$l_{\mathrm{D}}(t)=0$). Otherwise (when $l_{\mathrm{N}}(t)=0$, $l_{\mathrm{D}}(t)=1$
or $l_{\mathrm{N}}(t)=0$, $l_{\mathrm{D}}(t)=0$), the control input
at the plant side is set to $K_{\mathrm{N}}\tilde{x}(t)$ received
from the delay-free channel. Hence, the control input applied at the
plant is given by 
\begin{equation}
u(t)=(1-l_{\mathrm{N}}(t))K_{\mathrm{N}}\tilde{x}(t)+l_{\mathrm{N}}(t)(1-l_{\mathrm{D}}(t))K_{\mathrm{D}}\tilde{x}(t-1),\label{eq:delinput}
\end{equation}
 for $t\geq1$. Assuming $u(0)=0$, the closed-loop dynamics (\ref{eq:system}),
(\ref{eq:delinput}) can be given by 
\begin{align}
 & \left[\begin{array}{c}
\tilde{x}(t+2)\\
\tilde{x}(t+1)
\end{array}\right]\nonumber \\
 & \quad=\left[\begin{array}{cc}
A+(1-l_{\mathrm{N}}(t))BK_{\mathrm{N}} & l_{\mathrm{N}}(t)(1-l_{\mathrm{D}}(t))BK_{\mathrm{D}}\\
I_{\tilde{n}} & 0
\end{array}\right]\nonumber \\
 & \quad\quad\cdot\left[\begin{array}{c}
\tilde{x}(t+1)\\
\tilde{x}(t)
\end{array}\right],\quad t\in\mathbb{N}_{0}.\label{eq:delclosed}
\end{align}
By setting
\begin{align}
x(t)\triangleq\left[\begin{array}{c}
\tilde{x}(t+1)\\
\tilde{x}(t)
\end{array}\right],\,r(t) & \triangleq\begin{cases}
1, & l_{\mathrm{N}}(t)=0,\\
2, & l_{\mathrm{N}}(t)=1,\,l_{\mathrm{D}}(t)=0,\\
3, & l_{\mathrm{N}}(t)=1,\,l_{\mathrm{D}}(t)=1,
\end{cases}\label{eq:threemodemodesignal}
\end{align}
for $t\in\mathbb{N}_{0}$, the closed-loop dynamics (\ref{eq:delclosed})
forms a switched system (\ref{eq:switched-system}) with $3$ modes
represented by 
\begin{align}
A_{1} & \triangleq\left[\begin{array}{cc}
A+BK_{\mathrm{N}} & 0\\
I_{\tilde{n}} & 0
\end{array}\right],\nonumber \\
A_{2} & \triangleq\left[\begin{array}{cc}
A & BK_{\mathrm{D}}\\
I_{\tilde{n}} & 0
\end{array}\right],\quad A_{3}\triangleq\left[\begin{array}{cc}
A & 0\\
I_{\tilde{n}} & 0
\end{array}\right].\label{eq:threemodesa2a3}
\end{align}

Concerning the communication channels in the networked control system,
we assume the following. 

\begin{assumption} \label{AssumptionDelay} There exist scalars $\sigma_{\mathrm{N}},\sigma_{\mathrm{D}},\rho_{\mathrm{N}},\rho_{\mathrm{D}}\in[0,1]$
such that 
\begin{align}
 & \liminf_{k\to\infty}\frac{1}{k}\sum_{t=0}^{k-1}l_{\mathrm{N}}(t)\geq\sigma_{\mathrm{N}},\,\,\limsup_{k\to\infty}\frac{1}{k}\sum_{t=0}^{k-1}l_{\mathrm{N}}(t)\leq\rho_{\mathrm{N}},\label{eq:delfreeassumption}\\
 & \liminf_{k\to\infty}\frac{1}{k}\sum_{t=0}^{k-1}l_{\mathrm{D}}(t)\geq\sigma_{\mathrm{D}},\,\,\limsup_{k\to\infty}\frac{1}{k}\sum_{t=0}^{k-1}l_{\mathrm{D}}(t)\leq\rho_{\mathrm{D}},\label{eq:onestepdelassumption}
\end{align}
 hold almost surely. \end{assumption} 

Note that this assumption provides lower- and upper-bounds on the
long-run average numbers of transmission failures on the delay-free
and $1$-step-delayed channels. 

In the following result, we show that under Assumption~\ref{AssumptionDelay},
the mode signal of the switched system representing the networked
control system satisfies Assumption~\ref{MainAssumption-1}. 

\begin{proposition} \label{Proposition:rhofromsigmaandrho}Suppose
(\ref{eq:delfreeassumption}) and (\ref{eq:onestepdelassumption})
in Assumption~\ref{AssumptionDelay} are satisfied. Then (\ref{eq:infcond})
and (\ref{eq:supcond}) in Assumption~\ref{MainAssumption-1} hold
with 
\begin{align}
 & \underline{\rho}_{1}=1-\rho_{\mathrm{N}},\,\overline{\rho}_{1}=1-\sigma_{\mathrm{N}},\label{eq:firstres}\\
 & \underline{\rho}_{2}=\max\{0,\sigma_{\mathrm{N}}-\rho_{\mathrm{D}}\},\,\overline{\rho}_{2}=\min\{\rho_{\mathrm{N}},1-\sigma_{\mathrm{D}}\},\label{eq:secondres}\\
 & \underline{\rho}_{3}=\max\{0,\sigma_{\mathrm{N}}+\sigma_{\mathrm{D}}-1\},\,\overline{\rho}_{3}=\min\{\rho_{\mathrm{N}},\rho_{\mathrm{D}}\}.\label{eq:lastres}
\end{align}
\end{proposition}

\begin{IEEEproof} We use Lemma~\ref{LemmaXi} to show the result.
To this end, note that 
\begin{align}
\mathbbm{1}[r(t)=1] & =1-l_{\mathrm{N}}(t),\label{eq:firstr}\\
\mathbbm{1}[r(t)=2] & =l_{\mathrm{N}}(t)(1-l_{\mathrm{D}}(t)),\label{eq:secondr}\\
\mathbbm{1}[r(t)=3] & =l_{\mathrm{N}}(t)l_{\mathrm{D}}(t),\quad t\in\mathbb{N}_{0}.\label{eq:lastr}
\end{align}
First, we show (\ref{eq:infcond}) and (\ref{eq:supcond}) hold for
the case $s=1$ with $\underline{\rho}_{1},\overline{\rho}_{1}$ given
in (\ref{eq:firstres}). By (\ref{eq:delfreeassumption}) and (\ref{eq:onestepdelassumption}),
the inequalities (\ref{eq:xiliminf}) and (\ref{eq:xilimsup}) in
Lemma~\ref{LemmaXi} hold with $\xi_{1}(\cdot),\xi_{2}(\cdot)$ given
by $\xi_{1}(t)=1-l_{\mathrm{N}}(t)$, $\xi_{2}(t)=1$, $t\in\mathbb{N}_{0}$,
and $\varsigma_{1}=1-\rho_{\mathrm{N}}$, $\varrho_{1}=1-\sigma_{\mathrm{N}}$,
$\varsigma_{2}=1$, $\varrho_{2}=1$. By the lemma, we obtain (\ref{eq:infcond})
from (\ref{eq:xicombinedliminf}) and (\ref{eq:supcond}) from (\ref{eq:xicombinedlimsup})
with $\underline{\rho}_{1},\overline{\rho}_{1}$ given in (\ref{eq:firstres}).

Next, we show (\ref{eq:infcond}) and (\ref{eq:supcond}) hold for
the case $s=2$ with $\underline{\rho}_{2},\overline{\rho}_{2}$ given
in (\ref{eq:secondres}). By (\ref{eq:delfreeassumption}) and (\ref{eq:onestepdelassumption}),
the inequalities (\ref{eq:xiliminf}) and (\ref{eq:xilimsup}) hold
with $\xi_{1}(\cdot),\xi_{2}(\cdot)$ given by $\xi_{1}(t)=l_{\mathrm{N}}(t)$,
$\xi_{2}(t)=1-l_{\mathrm{D}}(t)$, $t\in\mathbb{N}_{0}$, and $\varsigma_{1}=\sigma_{\mathrm{N}}$,
$\varrho_{1}=\rho_{\mathrm{N}}$, $\varsigma_{2}=1-\rho_{\mathrm{D}}$,
$\varrho_{2}=1-\sigma_{\mathrm{D}}$. By applying Lemma~\ref{LemmaXi},
we obtain (\ref{eq:infcond}) from (\ref{eq:xicombinedliminf}) and
(\ref{eq:supcond}) from (\ref{eq:xicombinedlimsup}) with $\underline{\rho}_{2},\overline{\rho}_{2}$
given in (\ref{eq:secondres}). The result for the other case ($s=3$)
is obtained similarly by using Lemma~\ref{LemmaXi} together with
(\ref{eq:lastr}). \end{IEEEproof}

Proposition~\ref{Proposition:rhofromsigmaandrho} shows that the
networked control system (\ref{eq:system}), (\ref{eq:delinput})
with communication channels satisfying Assumption~\ref{AssumptionDelay}
can be represented by a switched system with a mode signal that satisfies
Assumption~\ref{MainAssumption-1}. As a result, Theorem~\ref{Stability-Theorem}
and the linear programming problems developed in Section~\ref{sec:Linear-Programming-Methods}
can be used for the stability analysis. 

\section{Numerical Examples \label{sec:Numerical-Example}}

In this section, we illustrate the efficacy of our results by investigating
stability properties of networked control systems discussed in Sections~\ref{subsec:Control-over-Delay-Free}
and \ref{subsec:Control-over-Delayed}. 

\subsubsection*{A) Example 1}

Consider the system (\ref{eq:system}) with 
\begin{align}
A= & \left[\begin{array}{cc}
1 & 0.1\\
-0.5 & 1.1
\end{array}\right],\quad B=\left[\begin{array}{c}
0.1\\
1.2
\end{array}\right].\label{eq:a-and-b}
\end{align}
In \cite{cetinkaya2016tac}, we explored stabilization of this system
over a network that faces random and malicious packet losses. There,
we proposed a linear state feedback controller with feedback gain
$K=\left[\begin{array}{cc}
-2.9012 & -0.9411\end{array}\right]$. By utilizing Theorem 3.5 of \cite{cetinkaya2016tac}, we see that
the closed-loop system is almost surely asymptotically stable whenever
the scalar $\rho$ identified in Assumption~\ref{MainAssumption}
is inside the range $[0,0.411]$. In the following we show that even
for strictly larger values of $\rho$, the closed-loop system remains
almost surely asymptotically stable. 

For investigating stability of the closed-loop system (\ref{eq:system}),
(\ref{eq:control-input}), we first characterize it as a switched
system (\ref{eq:switched-system}) with two modes represented with
$A_{1}=A+BK$ and $A_{2}=A$. Notice that for this switched system,
Assumption~\ref{MainAssumption} implies (\ref{eq:long-run-average-bound}),
and as a result, Assumption~\ref{MainAssumption-1} holds with $\underline{\rho}_{1}=1-\rho$,
$\overline{\rho}_{1}=1$, $\underline{\rho}_{2}=0$, and $\overline{\rho}_{2}=\rho$. 

\begin{figure}[t]
\centering\includegraphics[width=0.85\columnwidth]{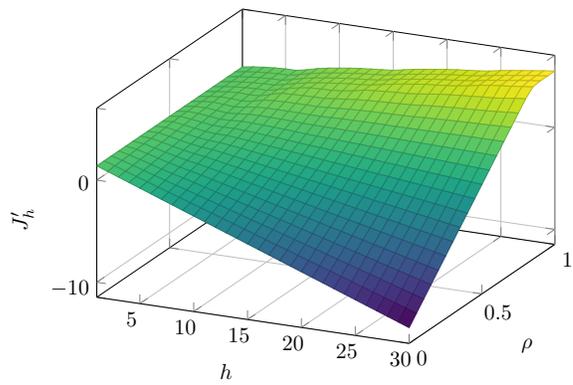} 

\caption{Optimal solution value $J'_{h}$ of the linear programming problem
(\ref{eq:linear-programming-problem-2}) with respect to $h$ and
$\rho$.}
 \label{Flo:jh1} 
\end{figure}

\begin{figure}[t]
\centering\includegraphics[width=0.85\columnwidth]{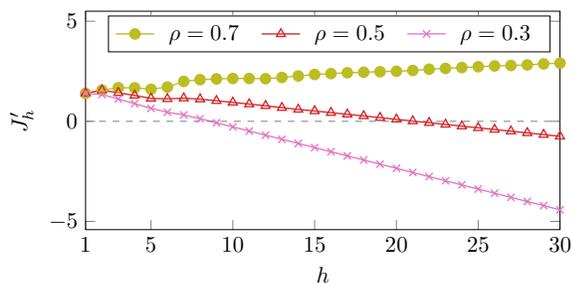} 

\caption{Optimal solution value $J'_{h}$ of the linear programming problem
(\ref{eq:linear-programming-problem-2}) with respect to $h$ for
$\rho=0.3$, $\rho=0.5$ and $\rho=0.7$.}

\label{Flo:jh2} 
\end{figure}

We numerically solve the linear programming problems (\ref{eq:linear-programming-problem-1})
and (\ref{eq:linear-programming-problem-2}) to obtain $J_{h}$ and
$J_{h}'$ for different values of $\rho$ and $h$. For finding the
coefficients $\gamma_{q}$ and $\gamma'_{z}$ of the objective functions,
we use (\ref{eq:small-gamma-definition}) (with the matrix norm induced
by the Euclidean norm and $\varepsilon=10^{-24}$) and (\ref{eq:lpp2-small-gamma-i-definition}).
We numerically confirm that $J_{h}=J'_{h}$ for $h=\{1,\ldots,11\}$.
For $h\geq12$, we utilize only the linear programming problem (\ref{eq:linear-programming-problem-2})
and obtain $J_{h}'$, as solving the problem (\ref{eq:linear-programming-problem-1})
takes excessively long times. We see in Fig.~\ref{Flo:jh1} that
for smaller values of $\rho$, $J_{h}'$ takes negative values indicating
almost sure asymptotic stability. In particular, we see in Fig.~\ref{Flo:jh2}
that when $\rho=0.5$, we obtain $J'_{22}<0$, which  implies that
(\ref{eq:maxcondition}) holds for all $\rho_{q}\in[0,1],q\in\mathcal{M}^{22}$,
that satisfy (\ref{eq:rhoqineq1}), (\ref{eq:rhoqineq2}). It follows
from Theorem~\ref{Stability-Theorem} that if Assumption~\ref{MainAssumption}
holds with $\rho=0.5$, and $\lim_{k\to\infty}\frac{1}{k}\sum_{t=0}^{k-1}\mathbbm{1}[\bar{l}(t)=q]$
exists for each $q\in\mathcal{M}^{22}$, then the zero solution of
the closed-loop system is almost surely asymptotically stable. We
note again that for $\rho=0.5$, the stability conditions of Theorem
3.5 in \cite{cetinkaya2016tac} do not hold. This indicates that the
stability conditions obtained in this paper are less conservative
than those in \cite{cetinkaya2016tac}. 

Note that for a given $\rho$, obtaining a nonnegative value for $J_{h}=J_{h}'$
does not necessarily imply that the system is unstable. For the same
$\rho$, the value of $J_{h}'$ may be positive for small $h$ and
negative for sufficiently large $h$ (see Remark~\ref{Remark-Jh-h-issue}).
For instance, in this example, $J'_{10}$ takes a negative value for
$\rho=0.3$, but not for $\rho=0.5$ or $\rho=0.7$ (Fig.~\ref{Flo:jh2}).
If one can only compute $J'_{h}$ up to $h=10$ due to limited computational
power, then stability for $\rho=0.5$ cannot be concluded. 

\begin{figure}[t]
\centering\includegraphics[width=0.85\columnwidth]{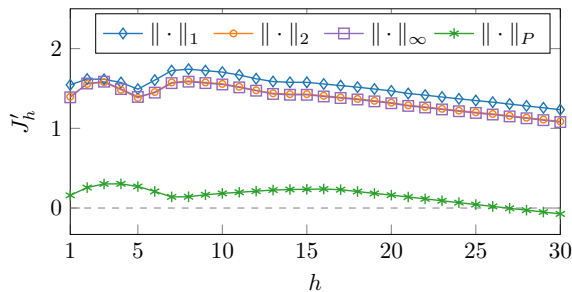} 

\caption{Optimal solution value $J'_{h}$ of the linear programming problem
(\ref{eq:linear-programming-problem-2}) with respect to $h$ for
fixed $\rho=0.6$ and different norms used in (\ref{eq:small-gamma-definition}).}

\label{Flo:example1_norm_comparison} 
\end{figure}

We also remark that using different matrix norms in the definition
of $\gamma_{q},q\in\mathcal{M}^{h}$, given in (\ref{eq:small-gamma-definition})
results in different trajectories for $J_{h}'$. For illustration,
we compute $J'_{h}$ with matrix norms $\|\cdot\|_{1}$, $\|\cdot\|_{2}$,
$\|\cdot\|_{\infty}$ (induced respectively by the $1$-norm, the
Euclidean norm, and the infinity norm of vectors), as well as the
matrix norm $\|\cdot\|_{P}$ induced by the vector norm $\|x\|_{P}\triangleq\sqrt{x^{\mathrm{T}}Px}$.
Here, we use the positive-definite matrix $P$ that we previously
utilized in \cite{cetinkaya2016tac} for the Lyapunov-based stability
analysis of this system. Fig. \ref{Flo:example1_norm_comparison}
shows the optimal solution value $J'_{h}$ of the linear programming
problem (\ref{eq:linear-programming-problem-2}) for different values
of $h$ when $\rho=0.6$. For this example, the values of $J'_{h}$
obtained with the matrix norm $\|\cdot\|_{P}$ is clearly lower than
others. We also observe that $J'_{30}$ obtained with $\|\cdot\|_{P}$
is negative. Thus, (\ref{eq:maxcondition}) holds for all $\rho_{q}\in[0,1],q\in\mathcal{M}^{30}$,
that satisfy (\ref{eq:rhoqineq1}), (\ref{eq:rhoqineq2}). It then
follows from Theorem~\ref{Stability-Theorem} if Assumption~\ref{MainAssumption}
holds with $\rho=0.6$, and $\lim_{k\to\infty}\frac{1}{k}\sum_{t=0}^{k-1}\mathbbm{1}[\bar{l}(t)=q]$
exists for each $q\in\mathcal{M}^{30}$, then the zero solution of
the closed-loop system is almost surely asymptotically stable. Hence,
for instance, the system is stable under all periodic attack scenarios
with $\rho\in[0,0.6]$, since Proposition~\ref{Proposition-l-g}
implies that the limits $\lim_{k\to\infty}\frac{1}{k}\sum_{t=0}^{k-1}\mathbbm{1}[\bar{l}(t)=q]$,
$q\in\mathcal{M}^{30}$, exist in the periodic case. 

In fact this stability region $[0,0.6]$ is quite tight, as there
exists a \emph{destabilizing} attack strategy for $\rho=0.64$. This
attack strategy is periodic with a period of $150$ time steps. It
was identified through the solution $\rho_{z},z\in\mathcal{Z}_{30}$,
to the linear programming problem (\ref{eq:linear-programming-problem-2})
solved with $h=30$ by using the matrix norm $\|\cdot\|_{2}$ and
the bounds $\underline{\rho}_{1}=1-\rho$, $\overline{\rho}_{1}=1$,
$\underline{\rho}_{2}=0$, $\overline{\rho}_{2}=\rho$, where $\rho=0.64$.
Under this attack strategy, the mode signal of the associated switched
system (\ref{eq:switched-system}) repeats the same pattern in every
$150$ time steps. This pattern is composed of a particular sequence
$\tilde{q}\in\mathcal{M}^{30}$ appearing once and then another sequence
$\hat{q}\in\mathcal{M}^{30}$ appearing $4$ times. Note that under
this attack strategy, Assumption~\ref{MainAssumption} holds with
$\rho=0.64$, and furthermore, the monodromy matrix associated with
the closed-loop periodic networked control system possesses an eigenvalue
that is outside the unit circle of the complex plane indicating divergence
of the state. 

\begin{nopf}

Under this attack strategy, in every $150$ time steps, exchange failure
process $l(\cdot)$ repeats the same pattern, which is composed of
the sequence $(0,0,0,1,1,1,1,1,1,0,0,0,1,1,1,1,1,1,0,0,0,1,1,1,1,1,1,0,0,0)$
appearing once and then the sequence $(0,1,1,1,1,1,1,0,1,1,1,1,1,1,0,0,1,1,1,1,1,1,0,1,1,1,1,1,1,0)$
appearing $4$ times.

\end{nopf}

\subsubsection*{B) Example 2}

In this example we demonstrate the results for the networked control
setup discussed in Section~\ref{subsec:Control-over-Delayed}. Specifically,
we consider the plant with $A$ and $B$ given by (\ref{eq:a-and-b})
in the previous subsection. The control packets are transmitted to
the plant over the delay-free and the $1$-step-delayed channels depicted
in Fig.~\ref{operation-2}. The feedback gains associated with these
channels are given by $K_{\mathrm{N}}=\left[\begin{array}{cc}
-2.9012 & -0.9411\end{array}\right]$ and $K_{\mathrm{D}}=[\begin{array}{cc}
-0.04 & -0.3\end{array}]$. We note that $K_{\mathrm{N}}$ is the gain from the previous subsection,
and $K_{\mathrm{D}}$ ensures that $A_{2}$ (of the equivalent switched
system formulation (\ref{eq:switched-system}) with (\ref{eq:threemodesa2a3}))
is a Schur matrix. 

\begin{figure}
\begin{center}\includegraphics[width=0.7\columnwidth]{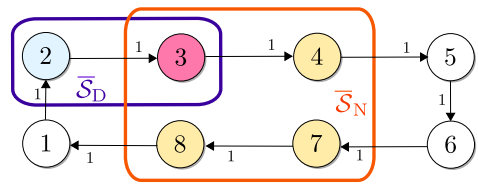}\end{center} 

\caption{Transition diagram for $\{g(t)\in\mathcal{S}\}_{t\in\mathbb{N}_{0}}$
representing an $8$-periodic jamming attack strategy against the
networked control system depicted in Fig~\ref{operation-2}. With
initial condition $g(0)=1$, the delay-free channel is periodically
attacked at times $8t+2$, $8t+3$, $8t+6$, $8t+7$, and moreover,
the $1$-step-delayed channel is periodically attacked at times $8t+1$,
$8t+2$, for $t\in\mathbb{N}_{0}$. }
 \label{Flo:Periodic-Chain-2} 
\end{figure}

We consider the case where the channels are subject to coordinated
periodic jamming attacks. In this case, $\{l_{\mathrm{N}}(t)\in\{0,1\}\}_{t\in\mathbb{N}_{0}}$
and $\{l_{\mathrm{D}}(t)\in\{0,1\}\}_{t\in\mathbb{N}_{0}}$, the failure
indicators of the delay-free and the one-step-delayed channels, can
be given by 
\begin{align*}
l_{\mathrm{N}}(t) & =\begin{cases}
1, & g(t)\in\overline{\mathcal{S}}_{\mathrm{N}},\\
0, & \mathrm{otherwise},
\end{cases}\,\,\,\,\,l_{\mathrm{D}}(t)=\begin{cases}
1, & g(t)\in\overline{\mathcal{S}}_{\mathrm{D}},\\
0 & \mathrm{otherwise},
\end{cases}
\end{align*}
where $\{g(t)\in\mathcal{S}\}_{t\in\mathbb{N}_{0}}$ is a finite-state
irreducible and periodic Markov chain with transition probabilities
either $0$ or $1$, and moreover, $\overline{\mathcal{S}}_{\mathrm{N}}$
and $\overline{\mathcal{S}}_{\mathrm{D}}$ are subsets of $\mathcal{S}$.
Fig.~\ref{Flo:Periodic-Chain-2} shows the transition diagram of
$\{g(t)\in\mathcal{S}\}_{t\in\mathbb{N}_{0}}$ with an $8$-periodic
pattern. At time $t$, the transmission on the delay-free channel
is attacked if $g(t)\in\overline{\mathcal{S}}_{\mathrm{N}}=\{3,4,7,8\}$;
moreover, the transmission on the $1$-step-delayed channel is attacked
if $g(t)\in\overline{\mathcal{S}}_{\mathrm{D}}=\{2,3\}$, and both
channels are attacked when $g(t)=3$. Note that for any attack strategy
represented with an irreducible $\{g(t)\in\mathcal{S}\}_{t\in\mathbb{N}_{0}}$,
the mode signal given by (\ref{eq:threemodemodesignal}) satisfies
(\ref{eq:l-g-definition}) with $\mathcal{S}_{1}\triangleq\mathcal{S}\setminus\overline{\mathcal{S}}_{\mathrm{N}}$,
$\mathcal{S}_{2}\triangleq\overline{\mathcal{S}}_{\mathrm{N}}\cap(\mathcal{S}\setminus\overline{\mathcal{S}}_{\mathrm{D}})$,
and $\mathcal{S}_{3}\triangleq\overline{\mathcal{S}}_{\mathrm{N}}\cap\overline{\mathcal{S}}_{\mathrm{D}}$.
Hence, it follows from Proposition~\ref{Proposition-l-g} that for
all $h\in\mathbb{N}$, the limits $\lim_{k\to\infty}\frac{1}{k}\sum_{i=0}^{k-1}\mathbbm{1}[\bar{r}(i)=q]$,
$q\in\mathcal{M}^{h}$, exist. 

To investigate the stability of the networked control system, we consider
different scenarios where the long-run-average transmission failures
on both channels satisfy Assumption~\ref{AssumptionDelay}. Then,
Proposition~\ref{Proposition:rhofromsigmaandrho} implies that the
switched system representation satisfies Assumption~\ref{MainAssumption-1}
with $\underline{\rho}_{s},\overline{\rho}_{s}$, $s\in\{1,2,3\}$,
given by (\ref{eq:firstres})--(\ref{eq:lastres}) as functions of
$\sigma_{\mathrm{N}},\sigma_{\mathrm{D}}$, $\rho_{\mathrm{N}},\rho_{\mathrm{D}}$. 

We first consider the case where the information about the average
number of failures on delay-free and $1$-step-delayed channels is
limited. In particular, we set the lower-bounds on the long-run average
number of failures on both channels to $0$, that is, $\sigma_{\mathrm{N}}=\sigma_{\mathrm{D}}=0$.
Our goal is to identify upper-bounds on the long-run average number
of failures of the channels ($\rho_{\mathrm{N}}$ and $\rho_{\mathrm{D}}$)
for which the closed-loop networked control system is stable. To this
end, we solve the linear programming problem (\ref{eq:linear-programming-problem-2})
for different values of $\rho_{\mathrm{N}}$ and $\rho_{\mathrm{D}}$.
For finding the coefficients $\gamma'_{z}=\max_{q\in\mathcal{M}^{h,z}}\gamma_{q},$
$z\in\mathcal{Z}_{h}$, of the objective function, we use (\ref{eq:small-gamma-definition})
(with the matrix norm induced by the Euclidean norm and $\varepsilon=10^{-24}$). 

\begin{figure}[t]
\centering\includegraphics[width=0.85\columnwidth]{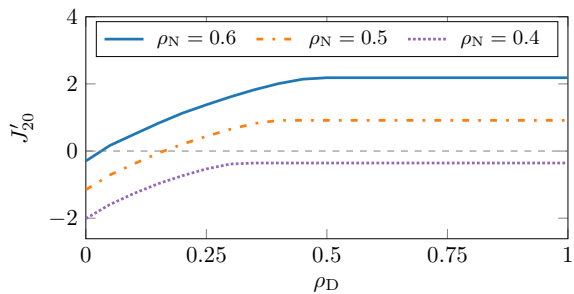} 

\caption{Optimal solution value $J'_{20}$ of the linear programming problem
(\ref{eq:linear-programming-problem-2}) with respect to $\rho_{\mathrm{D}}$
for the cases $\rho_{\mathrm{N}}=0.4$, $\rho_{\mathrm{N}}=0.5$,
and $\rho_{\mathrm{N}}=0.6$. (We set $\sigma_{\mathrm{D}}=\sigma_{\mathrm{N}}=0$
for all cases.)}

\label{Flo:example2-rho-comparison} 
\end{figure}

Fig.~\ref{Flo:example2-rho-comparison} shows how the optimal objective
function $J'_{20}$ of the linear programming problem (\ref{eq:linear-programming-problem-2})
changes with respect to $\rho_{\mathrm{D}}$ for the values $\rho_{\mathrm{N}}=0.4$,
$\rho_{\mathrm{N}}=0.5$, and $\rho_{\mathrm{N}}=0.6$. Observe that
when the long-run average number of failures on the delay-free communication
channel is sufficiently small, stability can be achieved regardless
of the amount of transmission failures on the $1$-step-delayed channel.
This is seen in Fig.~\ref{Flo:example2-rho-comparison} for the case
$\rho_{\mathrm{N}}=0.4$. Specifically, we have $J'_{20}<0$, which
implies that (\ref{eq:maxcondition}) holds for all $\rho_{q}\in[0,1],q\in\mathcal{M}^{20}$,
that satisfy (\ref{eq:rhoqineq1}), (\ref{eq:rhoqineq2}). Hence,
the stability can be concluded by Theorem~\ref{Stability-Theorem}
for all $\rho_{\mathrm{D}}\in[0,1]$. On the other hand, when the
delay-free channel faces more failures, the transmissions on the $1$-step-delayed
channel become more important. In particular for $\rho_{\mathrm{N}}=0.5$
and $\rho_{\mathrm{N}}=0.6$, the value of $J'_{20}$ is negative
and the stability can be concluded \emph{only} for sufficiently small
values of $\rho_{\mathrm{D}}$. 

\begin{figure}[t]
\centering\includegraphics[width=0.85\columnwidth]{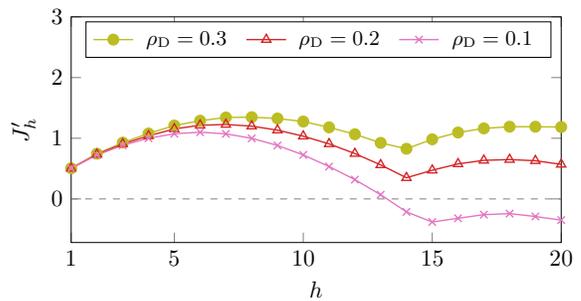} 

\caption{Optimal solution value $J'_{h}$ of the linear programming problem
(\ref{eq:linear-programming-problem-2}) with respect to $h$ for
the cases $\rho_{\mathrm{D}}=0.1$, $\rho_{\mathrm{D}}=0.2$, and
$\rho_{\mathrm{D}}=0.3$. (We set $\sigma_{\mathrm{D}}=0$, $\sigma_{\mathrm{N}}=\rho_{\mathrm{N}}=1$
for all cases.)}

\label{Flo:example-2-jh} 
\end{figure}

Next, we consider the situation where the delay-free channel is known
to be completely blocked due to jamming attacks at each time instant.
To explore this case, we set $\sigma_{\mathrm{N}}=\rho_{\mathrm{N}}=1$.
In this setup, all control packets are to be transmitted over the
$1$-step-delayed channel. We would like to find out the long-run
average number of failures that can be tolerated on this channel.
For this purpose, we solve the linear programming problem (\ref{eq:linear-programming-problem-2})
to obtain $J'_{h}$ with respect to $h$ for three different cases:
$\rho_{\mathrm{D}}=0.1$, $\rho_{\mathrm{D}}=0.2$, and $\rho_{\mathrm{D}}=0.3$.
In all cases we set $\sigma_{\mathrm{D}}=0$. Fig.~\ref{Flo:example-2-jh}
shows $J'_{h}$ for different values of $\rho_{\mathrm{D}}$. We observe
that $J'_{14}<0$ when $\rho_{\mathrm{D}}=0.1$, implying that (\ref{eq:maxcondition})
holds for all $\rho_{q}\in[0,1],q\in\mathcal{M}^{14}$, that satisfy
(\ref{eq:rhoqineq1}), (\ref{eq:rhoqineq2}). Hence, by Theorem~\ref{Stability-Theorem}
the zero solution of the closed-loop system is almost surely asymptotically
stable if the ratio of the transmission failures on the $1$-step-delayed
channel is bounded by $0.1$ in the long-run. On the other hand, for
$\rho_{\mathrm{D}}=0.2$ and $\rho_{\mathrm{D}}=0.3$, we cannot guarantee
stability, since we have $J'_{h}>0$, $h\in\{1,\ldots,20\}$. 

\begin{inarxiv}

\subsubsection*{C) Example 3}

Consider the system (\ref{eq:system}), (\ref{eq:control-input})
with 
\begin{align*}
A & =\left[\begin{array}{cc}
0 & 1\\
a_{1} & a_{2}
\end{array}\right],\,B=\left[\begin{array}{c}
0\\
1
\end{array}\right],\,K=\left[\begin{array}{cc}
-a_{1} & -a_{2}\end{array}\right],
\end{align*}
 where $a_{1}=2$, $a_{2}=1$. Suppose that the failure indicator
process $\{l(t)\in\{0,1\}\}_{t\in\mathbb{N}_{0}}$ satisfies Assumption~\ref{MainAssumption},
where $\rho$ denotes the probabilistic bound on the average ratio
of failures. This networked control system can be written as a switched
system (\ref{eq:switched-system}) with 
\begin{align*}
 & A_{1}=A+BK=\left[\begin{array}{cc}
0 & 1\\
0 & 0
\end{array}\right],\quad A_{2}=A=\left[\begin{array}{cc}
0 & 1\\
2 & 1
\end{array}\right].
\end{align*}
Following the formulation in Section~\ref{subsec:Control-over-Delay-Free},
let $\underline{\rho}_{1}\triangleq1-\rho,\overline{\rho}_{1}\triangleq1,\underline{\rho}_{2}\triangleq0$,
and $\overline{\rho}_{2}\triangleq\rho$. It follows from (\ref{eq:long-run-average-bound})
that Assumption~\ref{MainAssumption-1} is satisfied. Our goal is
to check the stability of this system by using Theorem~\ref{Stability-Theorem}.
To this end we utilize the linear programming problem (\ref{eq:linear-programming-problem-1}).
Notice that for this example, with $h=2$, we have 
\begin{align*}
\Gamma_{(1,1)} & =A_{1}^{2}=0
\end{align*}
 in (\ref{eq:lifted-system}). Hence, $\gamma_{(1,1)}$, in (\ref{eq:small-gamma-definition})
depends on the particular value selected for $\varepsilon$. With
the matrix norm induced by the Euclidean norm and $\varepsilon=10^{-24}$,
we solve the linear programming problem (\ref{eq:linear-programming-problem-1})
and obtain $J_{2}>0$ for every $\rho\geq0.5$. For instance, we have
$J_{2}=0.8047$ for $\rho=0.5$, $J_{2}=0.9328$ for $\rho=0.6$,
and $J_{2}=1.0609$ for $\rho=0.7$. Notice that when $\rho\geq0.5$,
either one of the switching sequences $2,1,2,1,\ldots$ or $1,2,1,2,\ldots$
are allowed. These sequences destabilize the system. As a result $J_{2}$
cannot be negative when $\rho\geq0.5$. In particular, when $\rho=0.5$,
an optimal solution is given by $\rho_{(2,1)}=1.0$ and $\rho_{q}=0$
for $q\neq(2,1)$, which corresponds to the switching sequence $2,1,2,1,\ldots$.
Notice that when $\rho>0.5$, there are also other destabilizing sequences
where the unstable mode (with subsystem matrix $A_{2}$) runs most
of the time. 

On the other hand, for $\rho<0.5$, $J_{2}$ can be obtained negative,
since in all optimal solutions, we have $\rho_{(1,1)}>0$. Notice
that the sequence $(1,1)$ indicates the pattern where the stable
mode (with subsystem matrix $A_{1}$) is active at consecutive time
instants. For example, when $\rho=0.48$, with $\varepsilon=10^{-16}$,
we obtain $J_{2}=-0.7011$, indicating stability. Notice that when
$\rho=0.49$, with $\varepsilon=10^{-16}$, we obtain a positive $J_{2}$
value, but by picking a smaller $\varepsilon=10^{-24}$, we get $J_{2}=-0.3166$
indicating stability. 

\end{inarxiv}

\section{Conclusion \label{sec:Conclusion}}

We explored almost sure asymptotic stability of a stochastic switched
linear system. Our proposed stability analysis approach relies on
studying the switched system's state at every $h\in\mathbb{N}$ steps.
We obtained sufficient stability conditions and showed that the stability
can be checked by solving a linear programming problem. The number
of variables in this problem grows polynomially in the number of subsystems,
but exponentially in $h$, which makes the computation difficult when
$h$ is large. To overcome this issue, we constructed an alternative
linear programming problem, where the number of variables grows polynomially
in both the number of subsystems and $h$. Even though the calculation
of the coefficients in the alternative problem takes additional time,
the solution is obtained faster compared to the original problem. 

Our linear programming-based analysis approach allows us to check
stability without relying on statistical information on the mode signal.
In particular, the probability of mode switches and the stationary
distributions associated with the modes are not needed for stability
analysis. We applied our approach in exploring networked control systems
under malicious jamming attacks. The technical challenge there is
that the attackers' specific strategies are not available for analysis.
By using our approach, we showed that stability can be guaranteed
under all possible attack strategies when the long-run average number
of network transmission failures satisfies certain conditions. In
practice, our approach can be used by system operators to assess the
safety of industrial processes. Specifically, our stability results
can be utilized in identifying the level of jamming attacks that can
be tolerated on the communication channels used for the measurement
and the control of a plant. 

Investigations of the case with noisy dynamics and the stabilization
problem are part of our future extensions. In the stabilization problem,
the controller may not have access to precise information of the active
mode. In those cases, the system modes are divided into several groups,
and the controller only knows which group contains the currently active
mode. This problem was considered for discrete- and continuous-time
Markov jump systems by \cite{do2002h,li2006stabilization}. For this
problem setting, our approaches may be extended for the case where
the mode signal is not necessarily a Markov process. 

\bibliographystyle{ieeetr}
\bibliography{references}

\appendix 

The following result provides lower- and upper-bounds for the long-run
average of the product of two binary-valued processes. 

\begin{aplemma} \label{LemmaXi} For all binary-valued processes
$\{\xi_{1}(t)\in\{0,1\}\}_{t\in\mathbb{N}_{0}}$ and $\{\xi_{2}(t)\in\{0,1\}\}_{t\in\mathbb{N}_{0}}$
that satisfy 
\begin{align}
 & \liminf_{k\to\infty}\frac{1}{k}\sum_{t=0}^{k-1}\xi_{i}(t)\geq\varsigma_{i},\label{eq:xiliminf}\\
 & \limsup_{k\to\infty}\frac{1}{k}\sum_{t=0}^{k-1}\xi_{i}(t)\leq\varrho_{i},\quad i\in\{1,2\},\label{eq:xilimsup}
\end{align}
 almost surely with $\varsigma_{i},\varrho_{i}\in[0,1]$, $i\in\{1,2\}$,
we have 
\begin{align}
\liminf_{k\to\infty}\frac{1}{k}\sum_{t=0}^{k-1}\xi_{1}(t)\xi_{2}(t) & \geq\max\{0,\varsigma_{1}+\varsigma_{2}-1\},\label{eq:xicombinedliminf}\\
\limsup_{k\to\infty}\frac{1}{k}\sum_{t=0}^{k-1}\xi_{1}(t)\xi_{2}(t) & \leq\min\{\varrho_{1},\varrho_{2}\},\label{eq:xicombinedlimsup}
\end{align}
 almost surely. 

\end{aplemma}

\begin{IEEEproof} To show (\ref{eq:xicombinedliminf}), first note
that 
\begin{align}
\liminf_{k\to\infty}\frac{1}{k}\sum_{t=0}^{k-1}\xi_{1}(t)\xi_{2}(t) & =1-\limsup_{k\to\infty}\frac{1}{k}\sum_{t=0}^{k-1}\left(1-\xi_{1}(t)\xi_{2}(t)\right).\label{eq:liminiflimsuprelation}
\end{align}
For all $i,j\in\{1,2\}$, $i\neq j$, we have 
\begin{align}
 & \limsup_{k\to\infty}\frac{1}{k}\sum_{t=0}^{k-1}\left(1-\xi_{1}(t)\xi_{2}(t)\right)\nonumber \\
 & \quad=\limsup_{k\to\infty}\frac{1}{k}\sum_{t=0}^{k-1}\left(\xi_{i}(t)(1-\xi_{j}(t))+\left(1-\xi_{i}(t)\right)\right)\nonumber \\
 & \quad\leq\limsup_{k\to\infty}\frac{1}{k}\sum_{t=0}^{k-1}\left(\xi_{i}(t)(1-\xi_{j}(t))\right)\nonumber \\
 & \quad\quad+\limsup_{k\to\infty}\frac{1}{k}\sum_{t=0}^{k-1}\left(1-\xi_{i}(t)\right).\label{eq:limsupsum}
\end{align}
 Since $\xi_{i}(t)(1-\xi_{j}(t))\leq\xi_{i}(t)$ and $\xi_{i}(t)(1-\xi_{j}(t))\leq1-\xi_{j}(t)$,
by (\ref{eq:xilimsup}), we have 
\begin{align*}
 & \limsup_{k\to\infty}\frac{1}{k}\sum_{t=0}^{k-1}\left(\xi_{i}(t)(1-\xi_{j}(t))\right)\leq\limsup_{k\to\infty}\frac{1}{k}\sum_{t=0}^{k-1}\xi_{i}(t)\leq\varrho_{i},
\end{align*}
and by (\ref{eq:xiliminf}), we have
\begin{align*}
 & \limsup_{k\to\infty}\frac{1}{k}\sum_{t=0}^{k-1}\left(\xi_{i}(t)(1-\xi_{j}(t))\right)\leq\limsup_{k\to\infty}\frac{1}{k}\sum_{t=0}^{k-1}\left(1-\xi_{j}(t)\right)\\
 & \quad\quad=1-\liminf_{k\to\infty}\frac{1}{k}\sum_{t=0}^{k-1}\xi_{j}(t)\leq1-\varsigma_{j}.
\end{align*}
Therefore, 
\begin{align}
\limsup_{k\to\infty}\frac{1}{k}\sum_{t=0}^{k-1}\left(\xi_{i}(t)(1-\xi_{j}(t))\right) & \leq\min\{\varrho_{i},1-\varsigma_{j}\}.\label{eq:firstxiresult}
\end{align}
 Furthermore, since $\limsup_{k\to\infty}\frac{1}{k}\sum_{t=0}^{k-1}\left(1-\xi_{i}(t)\right)\leq1-\varsigma_{i}$,
it follows from (\ref{eq:liminiflimsuprelation})--(\ref{eq:firstxiresult})
that 
\begin{align}
 & \liminf_{k\to\infty}\frac{1}{k}\sum_{t=0}^{k-1}\xi_{1}(t)\xi_{2}(t)\geq1-\left(\min\{\varrho_{i},1-\varsigma_{j}\}+1-\varsigma_{i}\right)\nonumber \\
 & \,\,=\varsigma_{i}-\min\{\varrho_{i},1-\varsigma_{j}\}=\max\{\varsigma_{i}-\varrho_{i},\varsigma_{i}+\varsigma_{j}-1\}.\label{eq:liminfineq1}
\end{align}
Now, noting that $\varsigma_{i}-\varrho_{i}\leq0$, $i\in\{1,2\}$,
and $\liminf_{k\to\infty}\frac{1}{k}\sum_{t=0}^{k-1}\xi_{1}(t)\xi_{2}(t)\geq0$,
almost surely, we have (\ref{eq:xicombinedliminf}) from (\ref{eq:liminfineq1}). 

Next, we prove (\ref{eq:xicombinedlimsup}). Since $\xi_{1}(t)\xi_{2}(t)\leq\xi_{1}(t)$
and $\xi_{1}(t)\xi_{2}(t)\leq\xi_{2}(t)$, we obtain 
\begin{align}
\limsup_{k\to\infty}\frac{1}{k}\sum_{t=0}^{k-1}\xi_{1}(t)\xi_{2}(t) & \leq\limsup_{k\to\infty}\frac{1}{k}\sum_{t=0}^{k-1}\xi_{i}(t),\label{eq:limsumineq1}
\end{align}
 for $i\in\{1,2\}$. It then follows from (\ref{eq:xilimsup}) that
$\limsup_{k\to\infty}\frac{1}{k}\sum_{t=0}^{k-1}\xi_{1}(t)\xi_{2}(t)\leq\varrho_{i}$,
$i\in\{1,2\}$, almost surely, which then implies (\ref{eq:xicombinedlimsup}).
\end{IEEEproof} 
\end{document}